\documentclass[journal]{IEEEtran}
\usepackage{siunitx}
\usepackage{booktabs}
\usepackage{graphicx}
\usepackage{caption}
\usepackage{subcaption}
\usepackage{comment}
\usepackage{datetime}
\usepackage{diagbox}

\usepackage{amsmath}
\usepackage{pgfplots}
\pgfplotsset{compat=1.16}
\usetikzlibrary{decorations}
\usetikzlibrary{decorations.markings}

\usepackage{xcolor}
\usepackage{hyperref}
\usepackage[belowskip=-12pt,aboveskip=12pt,font=footnotesize]{caption}
\usepackage[utf8]{inputenc}
\usepackage[ruled,vlined]{algorithm2e}
\usepackage{amsthm}
\DeclareMathOperator\erfc{erfc}
\newcommand{\numberthis}{\addtocounter{equation}{1}\tag{\theequation}}

\newtheorem{theorem}{Theorem}

\newtheorem{proposition}[theorem]{Proposition}
\newtheorem{corollary}[theorem]{Corollary}

\title{Channel Modeling for Multi-Receiver Molecular Communication Systems}
\author{\vspace{-.25cm}
Gokberk~Yaylali$^{\href{https://orcid.org/0000-0003-2582-9520}{\includegraphics[width=5pt]{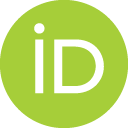}}}$,~\IEEEmembership{Student~Member,~IEEE}, Bayram~Cevdet Akdeniz$^{\href{https://orcid.org/0000-0002-9493-3105}{\includegraphics[width=5pt]{Figures/Orcidlogo.png}}}$,~\IEEEmembership{Member,~IEEE,} Tuna~Tugcu$^{\href{https://orcid.org/0000-0002-1332-3920}{\includegraphics[width=5pt]{Figures/Orcidlogo.png}}}$,~\IEEEmembership{Senior~Member,~IEEE,} and~Ali~Emre~Pusane$^{\href{https://orcid.org/0000-0002-8412-6684}{\includegraphics[width=5pt]{Figures/Orcidlogo.png}}}$,~\IEEEmembership{Senior~Member,~IEEE}%
\thanks{This work was supported in part by the Scientific and Technical Research Council of Turkey (TUBITAK) under Grant 119E190.}%
\thanks{G. Yaylali was with the Department of Electrical and Electronics Engineering, Bogazici University, 34342 Istanbul, Turkey. He is now with the Department of Electrical Engineering, Yale University, 06520 New Haven, CT, USA (e-mail: gokberk.yaylali@yale.edu).}%
\thanks{B. C. Akdeniz is with the Centre for Bioinformatics, University of Oslo, 0373 Oslo, Norway (e-mail: b.c.akdeniz@mn.uio.no).}%
\thanks{T. Tugcu is with NETLAB, Department of Computer Engineering, Bogazici University, 34342 Istanbul, Turkey (e-mail: tugcu@boun.edu.tr).}%
\thanks{A. E. Pusane is with the Department of Electrical and Electronics Engineering, Bogazici University, 34342 Istanbul, Turkey (e-mail: ali.pusane@boun.edu.tr).}%
\vspace{-.75cm}
}

\date{\today}

\begin{document}

\nocite{nakano2013molcom,yilmaz2017mcvd,kuran2011type,kabir2015quantity,garralda2011temporal,akdeniz2018position,gursoy2019index,srinivas2012wiener,nakano2012channel,yilmaz2014channel,zoofaghari2021semianalytical,zoofaghari2021biological,huang2020channel1d,bao2019channel,sabu2020far,yilmaz2020twoway,akdeniz2017optimal,tugcu2015receptor,akdeniz2018angular,Berezhkovskii1990mutualInfluence}

\maketitle
\setcounter{page}{1}
\begin{abstract}
Molecular Communication via Diffusion (MCvD) is a prominent small-scale technology, which roots from the nature. With solid analytical foundations on channel response and advanced modulation techniques, molecular single-input-single-output (SISO) systems are one of the most studied molecular networks in the literature. However, the literature is yet to provide sufficient analytical channel modeling on molecular multiple-output systems with fully absorbing receivers, {one of the common applications in the area. In this paper, a channel model for molecular single-input-multiple-output (SIMO) systems is proposed for estimating the channel response of such systems. With the model's recursive nature, the closed-form solution of the channel response of molecular 2-Rx SIMO systems is analytically derived. A simplified model with lower complexity is also presented at a cost of slightly less accurate channel estimation. The models are extended to the molecular SIMO systems with more than two receivers. The performance of the methods are evaluated for several topologies with different parameters, and the accuracy of the model is verified by comparing to computer-simulated channel estimations in terms of quantitative error metrics such as root-mean-squared error. The performance of the simplified model is verified by the amount of deviation, indicating sufficient channel modeling performance with reduced computational power.}
\end{abstract}

\markboth{UNDER REVIEW -- IEEE TRANSACTIONS ON COMMUNICATIONS}
\IEEEpeerreviewmaketitle
\section{Introduction}
\label{section:Introduction}
\IEEEPARstart{M}{olecular} communication (MC) is a novel communication technology inspired by the nature \cite{nakano2013molcom}. It enables nanoscale machinery to communicate within small ranges under the laws of physical phenomena. One of the most prominent areas of MC is molecular communication via diffusion (MCvD) \cite{yilmaz2017mcvd}, which obeys the laws of diffusion. MCvD systems use messenger molecules to convey information through fluidic medium. One of the most basic topologies used in MCvD is the molecular single-input-single-output (SISO) model. This fundamental topology consists of a single transmitter (Tx) and a single receiver (Rx). {In addition to molecular SISO systems, there are also multiple-entity systems in which multiple Rx and/or Tx may coexist. Information is encoded in many aspects of MCvD systems, such as molecule type \cite{kuran2011type}, quantity \cite{kabir2015quantity}, temporal position \cite{garralda2011temporal} \cite{akdeniz2018position} in molecular SISO systems, and index of the Tx antenna used \cite{gursoy2019index} in molecular MIMO systems. These message-carrying molecules propagate through the diffusive medium obeying the Brownian motion laws, at which the Brownian motion is modeled as a Wiener process \cite{srinivas2012wiener}. Due to this physical phenomenon, arrival of the messenger molecules is spread over time. In a communication scenario, a significant portion of these molecules arrive later than designated time interval, causing interference with the following symbols.} Consecutive symbols leave significantly large number of molecules roaming freely through the medium for the upcoming symbols. This outcome of the diffusion phenomenon, namely inter-symbol interference (ISI), severely worsens the communication quality of MC systems \cite{nakano2012channel}.

Molecular SISO systems have been extensively studied in the MC realm \cite{kuran2011type}. Specifically, molecular SISO system with a point transmitter and a spherical receiver is well-examined, and the corresponding channel characteristics are analytically derived \cite{yilmaz2014channel}. Asserting the channel behavior, analytical channel response is widely used to enhance the reach of the molecular SISO communication systems. However, molecular SISO systems cannot quite meet the requirements of high data transmission rates of today's technology \cite{gursoy2019index}, {thus the need for advanced modulation techniques on multiple-entity networks is increased.} Moreover, the future of molecular realm requires nanonetworks, where multiple entities need to coexist. With the rise of these networks, problems, such as synchronization, localization, and communication system design, have arisen and become of interest. 

In order to combat these problems, the need for proper analytical models for such multiple-entity networks, rather than simulation-based approaches, has emerged. Although there are some recent works {focused on the derivation of the} channel response with multiple receivers \cite{zoofaghari2021semianalytical, zoofaghari2021biological,huang2020channel1d}, they lack significant practical aspects. {For example, they are either obtained by fitting an empirical formula with some restrictions such as the receivers having to have same radius \cite{sabu2020far, Berezhkovskii1990mutualInfluence}, or they provide expressions that are too complex to be tractable with again radii constraints and spatial parameter restrictions \cite{bao2019channel}, \cite{yilmaz2020twoway}. Furthermore, all these works focus on channels with two receivers and they do not explicitly mention how to extend their approaches to the channels with more than two receivers.} To sum up, to the best of our knowledge, there is not a general framework to obtain exact and analytically tractable model for MCvD channels with multiple absorbing receivers. As a motivating example of the benefit of having analytically tractable channel response expression, proper analytical derivations for molecular SISO channel characteristics are conducted \cite{yilmaz2014channel} and due to its tractable structure, channel parameters, such as the peak time \cite{yilmaz2014channel} or optimum reception delay \cite{akdeniz2017optimal}, can be easily obtained. However, the same approach does not hold for multiple absorbing receiver networks due to dependence between receivers \cite{sabu2020far}. Therefore, acquiring a tractable channel response of multi-entity systems  without simulation-based approaches is still an open problem.

{In this paper, we, initially, propose an analytical channel response derivation for a 2-Rx molecular single-input-multiple-output (SIMO) topology rooted from the theoretical analysis of molecular SISO systems \cite{yilmaz2014channel}. For molecular SISO, it is straightforward to derive the channel response analytically from an arbitrary release point towards an arbitrarily placed spherical absorbing Rx \cite{yilmaz2014channel}. The channel response of each absorbing receiver could be analytically derived from molecular SISO analytical expressions \cite{yilmaz2014channel} if there were no other Rx. Additionally, the propagation of the molecules through the diffusive medium is a continuous Wiener process and is modeled by step increments independent of each other \cite{srinivas2012wiener}. Based on the independence of the random steps of molecules, a significant observation is made, that is, the succeeding movement of any molecule arriving at an arbitrary point in space can be modeled as if it were released from that arbitrary point. Using this observation, the channel response of an Rx in a molecular SIMO system can be deduced by estimating the fraction of \textit{stolen} molecules over time and subtracting from the channel response of the intended Rx. {The term ``stolen molecules" refers to the molecules absorbed by bystander Rxs, instead of the targeted Rx.} The work conducted in this paper is based on this observation, and aims to provide an analytical expression for the channel response of both receivers. Furthermore, the work intends to generalize the proposed channel model for multiple-entity networks with an arbitrary number of receivers greater than two.}

Organization of this paper is as follows: In Section~\ref{section:SystemModel}, 2-Rx molecular SIMO topology considered in the scope of this paper is provided. Afterwards, in Section~\ref{section:ProposedModels}, construction of the analytical expression regarding molecular SIMO topology is presented. In Subsection~\ref{subsection:RecursiveModel}, the comprehensive analytical expression is provided and afterwards, in Subsection~\ref{subsection:ApproxModel}, the simplified approximation is derived and presented. Accuracy and performance of these analytical expressions are provided in Section~\ref{section:Performance}, and Section~\ref{section:Conclusion} concludes the paper.

\section{System Model}
\label{section:SystemModel}
\begin{figure}[ht]
    \centering
    \includegraphics[width = .8\linewidth]{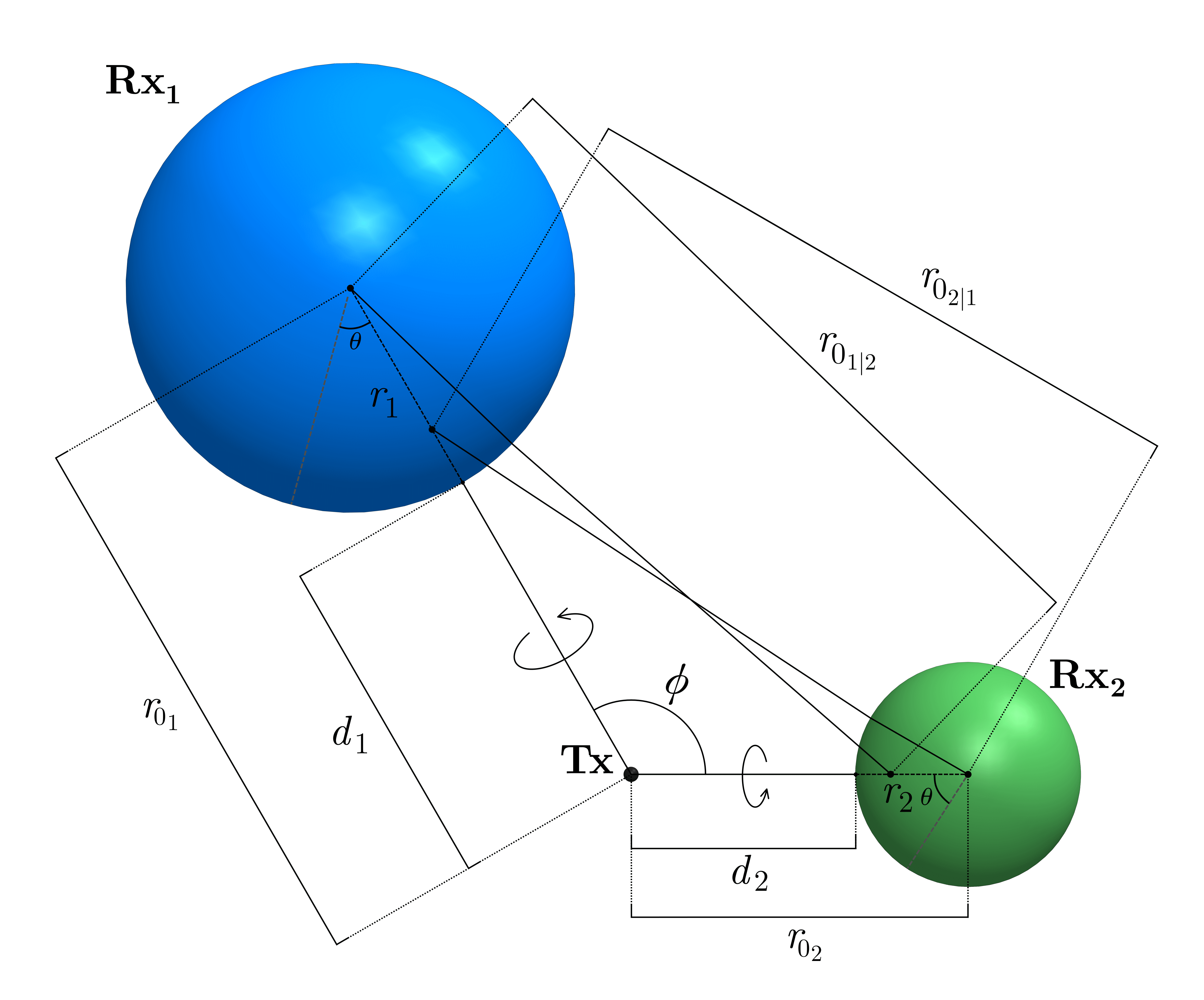}
    \caption{2-Rx molecular SIMO system model.}
    \label{fig:simo_model}
\end{figure}

In Fig.~\ref{fig:simo_model}, a 2-Rx molecular SIMO topology is presented. Tx is assumed to be infinitesimally small, {and is modeled as a point transmitter, which releases molecules instantly into the medium as impulses without initial velocity.} Environment is a boundless three-dimensional space, filled with fluid and exempt from drift. Propagation of the molecules are only subject to Brownian motion. Spherical receivers (Rx$_1$ and Rx$_2$) with arbitrary radii are placed at arbitrary center points in space. As shown in Fig.~\ref{fig:simo_model}, the radius of Rx$_i$ is $r_i$. Center-to-center distance between Tx and Rx$_i$ is $r_{0_i}$. The closest distance between Rx$_i$ and Tx is $d_i = r_{0_i} - r_i$. The receivers are assumed to absorb all molecules hitting their surfaces \cite{tugcu2015receptor}. Due to rotational symmetry formed around center-to-center axis between Tx and Rx (shown in figure), the distribution of the molecules hit on the surface of a spherical Rx depends only on the polar angle $\theta$ with respect to the symmetry axis, and does not depend on the azimuthal angle. The angular distribution of molecules arrived on Rx surface {as a function of $\theta$} is proposed in \cite{akdeniz2018angular}, and a fundamental observation is deduced that molecules tend to hit the front side of Rx facing Tx. The angular separation between the center-to-center axes of receivers $\phi$ is used to define the topology. {With the help of rotational symmetry, a point inside the Rx, which is closer to the closest point on the surface of Rx$_i$ to Tx, is approximated to be the center of the absorbed molecules, which will be explained in detail in Section~\ref{section:ProposedModels}.} {Evaluated to be used in the upcoming models, the distance between the center of the molecules absorbed by Rx$_i$ to Rx$_j$ is $r_{0_{j | i}}$.}

\begin{figure}[ht]
\centering
\begin{tikzpicture}
\begin{axis}[
    width=\linewidth, height=.7\linewidth, xlabel={Time $(s)$}, ylabel={Absorption probability}, ymin=0, ymax=0.35, xmin=0, xmax=5, grid,
    every axis plot/.append style={very thick},
    y tick label style={/pgf/number format/.cd, fixed, 
    fixed zerofill, precision=2 },
    legend style={at={(0.98,0.02)},anchor=south east,
    every axis plot/.append style={very thick},
    nodes={scale=0.65, transform shape}}]
\addplot[color=blue,dashed ] table[x=t,y=siso1] {data/comparison_data.txt};
\addplot[color=blue, postaction={ decoration={ markings, mark=between positions 0 and 1 step 0.1 with { \fill circle[radius=1.5pt]; },}, decorate,},forget plot] table[x=t,y=simres1] {data/comparison_data.txt};
\addlegendimage{color=blue, postaction={decoration={ markings, mark=between positions .5 and .5 step .5 with { \fill circle[radius=2pt]; },}, decorate,} }
\addplot[color=red,dashed] table[x=t,y=siso2] {data/comparison_data.txt};
\addplot[color=red, postaction={ decoration={ markings, mark=between positions 0 and 1 step 0.1 with { \fill circle[radius=1.5pt]; },}, decorate,},forget plot]  table[x=t,y=simres2] {data/comparison_data.txt};
\addlegendimage{color=red, postaction={decoration={ markings, mark=between positions .5 and .5 step .5 with { \fill circle[radius=2pt]; },}, decorate,} }
\legend{Molecular SISO distribution function of Rx$_1$ (Theoretical), Molecular SIMO distribution function of Rx$_1$ (Simulation), Molecular SISO distribution function of Rx$_2$ (Theoretical), Molecular SIMO distribution function of Rx$_2$ (Simulation)}    
\end{axis}
\end{tikzpicture}
\caption{Plots of absorption probability of molecules absorbed by Rx$_1$ and Rx$_2$ with $r_1 = \SI{6}{\micro\meter}$, $r_{0_1} = \SI{15}{\micro\meter}$, $r_2 = \SI{3}{\micro\meter}$, $r_{0_2}=\SI{9}{\micro\meter}$, placed with $\frac{2\pi}{3}\SI{}{\radian}$ angular separation. Theoretical distribution functions in a molecular SISO system without the presence of the other Rx (shown in blue and teal), and simulation-based distribution functions in a molecular SIMO system in presence of the other Rx (shown in red and orange), with $N = \SI{e6}{}$ molecules, $\Delta t = \SI{e-4}{\second}$.}
\label{fig:comparison_plot}
\end{figure}

As previously proposed in literature \cite{yilmaz2014channel}, the absorption rate of molecules to a receiver in a molecular SISO system $p_{\text{hit}}^{\text{SISO}}(\text{Rx}, t \vert\ r_{0},r_r)$ is
\begin{multline}
\label{eqn:siso_pdf}
p_{\text{hit}}^{\text{SISO}}(\text{Rx}, t \vert\ r_{0},r_r) = \\
\frac{r_r}{r_{0}} \frac{1}{\sqrt{4\pi D t}} \frac{r_{0} - r_r}{t} \exp \Bigg[-\frac{(r_{0}-r_r)^2}{4Dt} \Bigg].
\end{multline}
{where $r_r$ is the radius of Rx, and $D$ is the diffusion coefficient.} The absorption probability of molecules to a receiver until time $t$ is obtained by integrating (\ref{eqn:siso_pdf}) over time, as
\begin{multline}
\label{eqn:siso_cdf}
P_{\text{hit}}^{\text{SISO}}(\text{Rx}, t \vert\ r_{0},r_r) = \\ \int_0^t p_{\text{hit}}^{\text{SISO}}(\text{Rx}, \tau \vert\ r_{0},r_r) d\tau = \frac{r_r}{r_{0}} \erfc{}\left[\frac{r_0 - r_r}{\sqrt{4Dt}}\right],
\end{multline}
{where $\erfc[\cdot]$ is the complementary error function.} As shown in Figs.~\ref{fig:comparison_plot}, the existence of another receiver significantly affects the channel response of the intended receiver. In boundless space, molecules propagate towards all directions with random movement and form the molecular SISO absorption rate function for molecular SISO systems \cite{yilmaz2014channel}. However, for molecular SIMO systems, several messenger molecules are absorbed by the other receiver during their random movement, and these molecules are \textit{stolen} from the intended receiver, compared to the molecular SISO case. Due to this inevitable and reciprocal dependence between individual receivers, the channel response of each receiver in presence of the other one cannot be acquired directly via the molecular SISO analytics. Therefore, proper analytical expressions for multiple-receiver topologies are required.



\section{Proposed Models}
\label{section:ProposedModels}

In this section, we first propose a novel approach that utilizes the molecular SISO model in a recursive fashion to model molecular SIMO molecule reception. We, then, provide an accurate simplification of the proposed model. The channel impulse response of a system is acquired from the output of the system given that the input is an impulse signal. In molecular communication, messenger molecules are most often released as impulses. So, the absorption rate of the molecules towards Rx over time is a characteristic feature of the system. As mentioned above, the motion of the messenger molecules is modeled as a Wiener process \cite{srinivas2012wiener}, which has independent step increments. Thus, the motion of a molecule can be split into parts, and each part is independent of each other. In this regard, one can model the motion of molecules arriving at an arbitrary point, and the motion of those molecules starting their motion from that arbitrary point towards any other point in space, separately.
    
\subsection{Recursive Model}
\label{subsection:RecursiveModel}

Molecular SISO {channel model accurately models the channel response of a single receiver in a molecular SISO system}. However, the existence of another {absorbing} receiver interrupts {the singularity in the space}, and reduces the number of molecules arriving at the intended receiver. This effect is reciprocal, since each receiver interrupts the space with respect to the other one. In this respect, modeling the channel response of both receivers with analytical expressions of molecular SISO model causes an overcounting problem.

We can start with a molecular SIMO topology, similar to the one given in Fig.~\ref{fig:simo_model}, {where Rx$_i$ is the intended receiver in the presence of Rx$_j$, $i$ and $j$ being the differing receiver indices.}
\begin{proposition}
\label{prop:prop1}
The presence of other absorbing receiver, Rx$_j$, would not aid the absorption rate of the molecules to Rx$_i$.
\end{proposition}
Using the Proposition~\ref{prop:prop1}, one simple result can be deducted.
\begin{corollary}
\label{cor:cor1}
The upper bound of the molecular SIMO hitting density function of Rx$_i$ is molecular SISO hitting density function, without the presence of Rx$_j$.
\end{corollary}
The formulation of Corollary~\ref{cor:cor1} is
\begin{equation}
\label{eq:corollary2}
p_{\text{hit}}^{\text{SIMO}}(\text{Rx}_i,t \vert\ \text{Rx}_j, r_{0_i}, r_i) \leq p_{\text{hit}}^{\text{SISO}}(\text{Rx}_i, t \vert\ r_{0_i},r_i).
\end{equation}
Coexisting receivers compete to absorb the molecules. Theoretically, one receiver can steal as much as the amount in a molecular SIMO system, at most. As a result, the largest possible reduction in the absorption rate of molecules of Rx$_i$ in molecular SIMO system conducted by Rx$_j$ is $p_{\text{hit}}^{\text{SISO}}(\text{Rx}_j, t \vert r_{0_j},r_j)$, and vice versa. Thus, the lower and upper bounds of the hitting {rate of molecules of both receivers in total can be stated by}
\begin{multline}
\max \Big[p_{\text{hit}}^{\text{SISO}}(\text{Rx}_1, t \vert r_{0_1},r_1),\ p_{\text{hit}}^{\text{SISO}}(\text{Rx}_2, t \vert r_{0_2},r_2) \Big]\\
\leq p_{\text{hit}}^{\text{SIMO}}(\text{Rx}_1,t \vert\ \text{Rx}_2, r_{0_1}, r_1) + p_{\text{hit}}^{\text{SIMO}}(\text{Rx}_2,t \vert\ \text{Rx}_1, r_{0_2}, r_2)\\
\leq p_{\text{hit}}^{\text{SISO}}(\text{Rx}_1, t \vert\ r_{0_1},r_1) + p_{\text{hit}}^{\text{SISO}}(\text{Rx}_2, t \vert\ r_{0_2},r_2).
\end{multline}
For simplification purposes, the molecular SISO absorption rate function of Rx$_i$ $p_{\text{hit}}^{\text{SISO}}(\text{Rx}_i, t \vert r_{0_i},r_i)$ and the molecular SIMO absorption rate function of Rx$_i$ $p_{\text{hit}}^{\text{SIMO}}(\text{Rx}_i,t \vert\ \text{Rx}_j, r_{0_i}, r_i)$ are represented with shorter versions $p_{\text{hit}}^{\text{SISO}}(\text{Rx}_i, t )$, and $p_{\text{hit}}^{\text{SIMO}}(\text{Rx}_i,t)$ from now on, assuming the knowledge of the spatial parameters.

While trying to acquire the absorption rate of molecules for Rx$_i$, it is observed that a portion of the molecules released is absorbed by Rx$_j$. Thanks to the independence of the increments of the motion, the absorption rate of the molecules, that are absorbed by Rx$_j$, towards Rx$_i$ can be modeled independently as if they were not absorbed by Rx$_j$, but released from the absorption point on Rx$_j$. For a specific instance, the absorption rate of molecules that are absorbed by Rx$_j$ at time $\tau$ towards Rx$_i$ can be calculated as the absorption rate of molecules that are released at the absorption point on Rx$_j$ at time $\tau$ toward Rx$_i$. Notice that the absorption rate of molecules absorbed by Rx$_j$ at time $\tau$ is affected by the absorption rate of molecules absorbed by Rx$_i$ at time smaller than $\tau$, and so on. Therefore, the absorption rate of molecules for receivers in molecular SIMO system $p_{\text{hit}}^{\text{SIMO}}(\text{Rx}_i,t)$ need to be evaluated in a recursive manner, rather than in a molecular SISO manner.
\begin{figure}[ht]
    \centering
    \includegraphics[width = 0.7\linewidth]{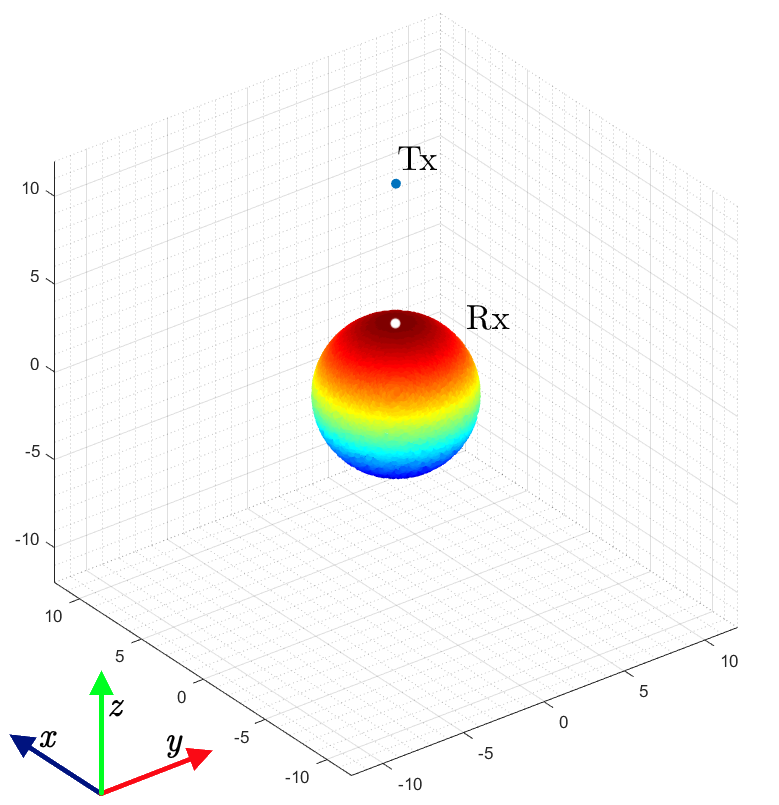}
    \caption{Heatmap of the distribution of molecules absorbed on a receiver in a molecular SISO system, generated via computer-aided diffusion simulation. Tx is placed at $(0,0,\SI{12}{\micro\meter})$ and center of Rx is placed in $(0,0,0)$. $r_r = \SI{4}{\micro\meter}$, $N=\SI{5e4}{}$, $D = \SI[per-mode=symbol]{79.4}{\micro\meter\squared \per \second}$, and $\Delta t = \SI{e-4}{\second}$. The closest point on Rx towards Tx is shown.}
    \label{fig:heatmap_siso}
\end{figure}
In this case, the positions of the molecules on Rx$_2$ surface become of importance. As stated in \cite{akdeniz2018angular}, the distribution of molecules on Rx surface depends only on the polar angle {for molecular SISO systems}, and the angular distribution of the absorption rate of molecules for a spherical absorbing receiver in a molecular SISO system as function of time $t$ \cite{akdeniz2018angular} is
\begin{multline}
\label{eq:siso_angular_dist}
p_{\text{hit}}^{\text{SISO}}(\text{Rx},t, \theta) = \\ \frac{ \sin{\theta} \erfc\left(\frac{r_0^*(\theta)}{\sqrt{4Dt}}\right)P_{\text{hit}}^{\text{SISO}}(\text{Rx},t)} {\left(1-\frac{2r_r}{r_0}\cos{\theta}+\frac{r_r^2}{r_0^2}\right)^{\frac{3}{2}} {\displaystyle \int_{0}^{\pi}} \frac{\sin{\theta'}\erfc\left(\frac{r_0^*(\theta')}{\sqrt{4Dt}}\right)} {\left(1-\frac{2r_r}{r_0}\cos{\theta'}+\frac{r_r^2}{r_0^2}\right)^{\frac{3}{2}}}d\theta'},
\end{multline}
where
\begin{equation}
r_0^*(\theta) = \sqrt{r_0^2 + r_r^2 -2r_0r_r\cos{\theta}}.
\end{equation}
As $t$ goes to infinity, the angular distribution of the absorption rate of molecules for Rx becomes \textit{saturated.} The marginal angular distribution of the absorption rate of the molecules \cite{akdeniz2018angular} becomes
\begin{align*}
p_{\text{hit}}^{\text{SISO}}(\text{Rx},\theta) &= 2\pi r_r^2 \sin{\theta} \frac{\left(1- \frac{r_r^2}{r_0^2} \right) }{4\pi r_r r_0 \left(1-\frac{2r_r}{r_0}\cos{\theta} + \frac{r_r^2}{r_0^2} \right)^{3/2} },\\
 &=  \frac{r_r \sin{\theta} \left(1- \frac{r_r^2}{r_0^2} \right) }{2r_0 \left(1-\frac{2r_r}{r_0}\cos{\theta} + \frac{r_r^2}{r_0^2} \right)^{3/2} }. \numberthis
\label{eq:siso_angular_dist_saturated}
\end{align*}
To calculate the center of mass of the molecules absorbed on Rx, one needs to calculate the expected value of $r_r \cos{\theta}$, which gives the expected value of the distance of the center of mass to the center of Rx, as
\begin{align*}
E\left[r_r \cos{\theta} \right] &= \int_0^\pi r_r \cos{\theta^\prime} \frac{p_{\text{hit}}^{\text{SISO}}(\text{Rx},\theta^\prime)}{P_{\text{hit}}^{\text{SISO}}(\text{Rx},t \to \infty)} d\theta^\prime,\\
&= \frac{r_r \left(1- \frac{r_r^2}{r_0^2} \right) }{2} \int_0^\pi  \frac{\cos{\theta^\prime} \sin{\theta^\prime}} {\left(1-\frac{2r_r}{r_0}\cos{\theta} + \frac{r_r^2}{r_0^2} \right)^{3/2} } d\theta^\prime,\\
&= r_r \frac{r_r}{r_0}. \numberthis
\label{eq:siso_center_of_mass}
\end{align*}

The expected value of the center of mass of the molecules absorbed is in front of the center, closer to Tx. Supported by both the analytical angular distribution and the simulation-based methods shown in Fig.~\ref{fig:heatmap_siso}, it is straightforward to state that molecules are absorbed on the closer-to-Tx side of Rx more. {The existence of another absorbing Rx reciprocally changes the angular distribution of molecules on the surface of Rx. However, although analytically challenging to show due to recursivity, this shift-impact is negligible. Therefore, the center of mass of the molecules absorbed on an Rx within a molecular SIMO topology is also assumed to be azimuthally symmetric.}

{As time goes to infinity, the center of mass of the absorbed molecules converges to (\ref{eq:siso_center_of_mass}). In this case, the motion of the molecules absorbed by Rx$_j$ can be split into two movements: one being the release at Tx, and the other being the release from the absorption point on Rx$_j$ \textit{as if} they were not absorbed. The latter part takes the \textit{stealing effect} of Rx$_j$ on Rx$_i$ into account. For the latter part, the center of mass of the molecules absorbed can be picked as \textit{the virtual release point}, as if they were released from that point. Based on this assumption, the distance from the center of Rx$_j$ towards Tx shown in (\ref{eq:siso_center_of_mass}) can be approximated as the released point. For example, the molecules absorbed by Rx$_j$ at time $\tau < t$ could be considered as if they were released at time $\tau$ from the virtual release point, and the absorption rate of those molecules is calculated from the Rx$_i$ perspective \cite{yilmaz2020twoway}.}

Following the assumption, {the conditional-absorption rate of the molecules absorbed by Rx$_j$ at time $\tau$ towards Rx$_1$ can be given as}
\begin{equation}
\label{eq:conditional}
p_{\text{hit}}(\text{Rx}_i,t \vert \text{Rx}_j, \tau) = 
\begin{cases}
p_{\text{hit}}^{\text{SISO}}(\text{Rx}_i,t-\tau \vert r_{0_{i \vert j}},r_i), & t > \tau,\\
0, & t \leq \tau,
\end{cases}
\end{equation}
{where $r_{0_{i\vert j}}$ is the point-to-center distance between the virtual release point of Rx$_j$ to Tx, and the center of Rx$_i$. Note that} two receivers' centers and one transmitter form a plane in 3-D space. Centered at Tx, angular separation between Rxs is defined as $\phi$ (shown in Fig.~\ref{fig:simo_model}) and it is used to calculate {$r_{0_{j\vert i}}$ and $r_{0_{i \vert j}}$} distances. These distances are calculated using {the law of cosines} as
{\begin{align}
r_{0_{j \vert i}} &= \sqrt{\left(r_{0_i}-r_i\frac{r_i}{r_{0_i}}\right)^2 + r_{0_j}^2 - 2\left(r_{0_i}-r_i\frac{r_i}{r_{0_i}}\right)r_{0_j}\cos{\phi} },\\
r_{0_{i \vert j}} &= \sqrt{\left(r_{0_j}-r_j\frac{r_j}{r_{0_j}}\right)^2 + r_{0_i}^2 - 2\left(r_{0_j}-r_j\frac{r_j}{r_{0_j}}\right)r_{0_i}\cos{\phi} }.
\end{align}}
Notice that the conditional-hitting {rate} depends on time $\tau$ and is shifted in time by the amount $\tau$. {The stealing effect is reciprocal, so the amount of molecules stolen by Rx$_i$ from Rx$_j$ until time $\tau$ needs to be taken into account. Therefore, the absorption rate of molecules absorbed by Rx$_j$ at time $\tau$ is yet to be determined and needs to be expressed as $p_{\text{hit}}^{\text{SIMO}}(\text{Rx}_j, \tau)$. Combining this expression with (\ref{eq:conditional}) gives the joint-absorption rate of molecules, as if those molecules were absorbed by Rx$_j$ at time $\tau$ and then released again from the virtual release point and absorbed by Rx$_i$ at time $t > \tau$, which corresponds to the stealing effect of the molecules absorbed by Rx$_j$ at time $\tau$ on Rx$_i$ at time $t$, as}
\begin{align*}
p_{\text{hit}}(\text{Rx}_i,t;\ \text{Rx}_j,\tau) &= p_{\text{hit}}^{\text{SIMO}}(\text{Rx}_j,\tau) p_{\text{hit}}(\text{Rx}_i,t \vert \text{Rx}_j, \tau),\\
&= p_{\text{hit}}^{\text{SIMO}}(\text{Rx}_j,\tau) p_{\text{hit}}^{\text{SISO}}(\text{Rx}_i,t-\tau \vert r_{0_{i \vert j}},r_i). \numberthis
\label{eq:joint_hit}
\end{align*}
Notice that this joint-hitting {rate} is zero for $t < \tau$, as the conditional-hitting {rate} in (\ref{eq:conditional}) dictates. {To take the effect of all molecules absorbed by Rx$_j$ and stolen from Rx$_i$ at all time instances $\tau < t$ into account, the absorption rate of the total of the molecules is calculated by integrating (\ref{eq:joint_hit}) over $\tau < t$, as
\begin{align*}
p_{\text{hit}}(\text{Rx}_i,t;\ \text{Rx}_j) &= \int_{0}^{t} p_{\text{hit}}^{\text{SIMO}}(\text{Rx}_j, \tau) p_{\text{hit}}(\text{Rx}_i,t \vert \text{Rx}_j, \tau) d\tau,\\
= \int_{0}^{t} p_{\text{hit}}^{\text{SIMO}}&(\text{Rx}_j, \tau) p_{\text{hit}}^{\text{SISO}}(\text{Rx}_i,t-\tau \vert r_{0_{i \vert j}},r_i) d\tau. \numberthis
\label{eq:total_hit}
\end{align*}}
This acquired hitting {rate} stands for {the absorption rate of the molecules absorbed by Rx$_j$ at any time $\tau < t$ towards Rx$_i$} at time $t$. As stated in the beginning, the nature of absorbing receiver design forbids a molecule to propagate further after absorption. Therefore, the hitting {rate} of such molecules need to be subtracted from the molecular SISO response of Rx$_i$, in order to {measure the adverse effect of the presence of Rx$_j$:}
\begin{align*}
p_{\text{hit}}^{\text{SIMO}}(\text{Rx}_i,t) &= p_{\text{hit}}^{\text{SISO}}(\text{Rx}_i,t) - p_{\text{hit}}(\text{Rx}_i,t;\ \text{Rx}_j), \numberthis \label{eq:simo_generic}\\
&= p_{\text{hit}}^{\text{SISO}}(\text{Rx}_i,t) - \numberthis 
\label{eq:simo_integral} \\ &\int_{0}^{t} p_{\text{hit}}^{\text{SIMO}}(\text{Rx}_j, \tau) p_{\text{hit}}^{\text{SISO}}(\text{Rx}_i,t-\tau \vert r_{0_{i \vert j}},r_i) d\tau.
\end{align*}

Notice that the absorption rate expression $p_{\text{hit}}^{\text{SIMO}}(\text{Rx}_i,t)$ also contains molecular SIMO expression of the absorption rate of molecules towards Rx$_j$. Consequently, the molecular SIMO hitting {rate} expression for receivers has a recursive {and mutually-feeding} nature. Although the recursive model gives reliable results (as will be demonstrated in the following sections), the recursive nature is a disadvantage and will be simplified next. {It can be seen that (\ref{eq:total_hit}) is a convolution integral, and can be written in form of
\begin{equation}
p_{\text{hit}}(\text{Rx}_i,t;\ \text{Rx}_j) = p_{\text{hit}}^{\text{SIMO}}(\text{Rx}_j, t) \ast p_{\text{hit}}^{\text{SISO}}(\text{Rx}_i, t \vert r_{0_{i \vert j}},r_i),
\end{equation}
which makes (\ref{eq:simo_integral}) a simpler form as
\begin{align*}
p_{\text{hit}}^{\text{SIMO}}(\text{Rx}_i,t) =&p_{\text{hit}}^{\text{SISO}}(\text{Rx}_i,t) - \numberthis \label{eq:simo_convolution}\\
&p_{\text{hit}}^{\text{SIMO}}(\text{Rx}_j, t) \ast p_{\text{hit}}^{\text{SISO}}(\text{Rx}_i, t \vert r_{0_{i \vert j}},r_i). 
\end{align*}
{To derive the closed-form of the molecular SIMO absorption rate expression, (\ref{eq:simo_convolution}) and its reciprocal are transformed into Laplace domain as
\begin{align}
P_i^{\text{SIMO}}(s) &= P_i^{\text{SISO}}(s) - P_j^{\text{SIMO}}(s)P_{i\vert j}^{\text{SISO}}(s),\\
P_j^{\text{SIMO}}(s) &= P_j^{\text{SISO}}(s) - P_i^{\text{SIMO}}(s)P_{j\vert i}^{\text{SISO}}(s),
\end{align}
where $P_i^{\text{SIMO}}(s) = \mathcal{L}\big\{ p_{\text{hit}}^{\text{SIMO}}(\text{Rx}_i,t) \big\}$, $P_j^{\text{SIMO}}(s) = \mathcal{L}\big\{ p_{\text{hit}}^{\text{SIMO}}(\text{Rx}_j,t) \big\}$, $P_i^{\text{SISO}}(s) = \mathcal{L}\big\{p_{\text{hit}}^{\text{SISO}}(\text{Rx}_i,t) \big\}$, $P_j^{\text{SISO}}(s) = \mathcal{L}\big\{p_{\text{hit}}^{\text{SISO}}(\text{Rx}_j,t) \big\}$, $P_{i\vert j}^{\text{SISO}}(s) = \mathcal{L}\big\{ p_{\text{hit}}^{\text{SISO}}(\text{Rx}_i, t \vert r_{0_{i \vert j}},r_i) \big\}$, and $P_{j\vert i}^{\text{SISO}}(s) = \mathcal{L}\big\{ p_{\text{hit}}^{\text{SISO}}(\text{Rx}_j, t \vert r_{0_{j \vert i}},r_j) \big\}$. Substituting equations into each other utilizes the SIMO-independent Laplace expression as}
{\begin{equation}
P_i^{\text{SIMO}}(s) = \frac{P_i^{\text{SISO}}(s) - P_j^{\text{SISO}}(s)P_{i\vert j}^{\text{SISO}}(s)} {1-P_{i\vert j}^{\text{SISO}}(s) sP_{j\vert i}^{\text{SISO}}(s)}.
\end{equation}}
All remaining expressions are Laplace domain duals of the molecular SISO hitting {rate} expressions. Therefore, the remaining parts obey the same function form, presented in (\ref{eqn:siso_pdf}). Laplace transform of (\ref{eqn:siso_pdf}) is
\begin{equation}
\mathcal{L}\left\{ p_{\text{hit}}^{\text{SISO}}(\text{Rx}_i, t) \right\} = \frac{r_i}{r_{0_i}} \exp \left( - \frac{r_{0_i}-r_i}{\sqrt{D}} \sqrt{s} \right).
\end{equation}
Placing appropriate transforms of relevant parts into the expression gives
\begin{multline}
\label{eq:simo_laplace_expanded}
P_i^{\text{SIMO}}(s) = \frac{ \frac{r_i}{r_{0_i}} \exp\left(-\frac{r_{0_i}-r_i}{\sqrt{D}} \sqrt{s} \right) }{ 1-  \frac{r_i r_j}{r_{0_{i\vert j}} r_{0_{j\vert i}}} \exp\left(-\frac{\left( r_{0_{i\vert j}} + r_{0_{j\vert i}} \right)- \left( r_i+r_j \right)}{\sqrt{D}} \sqrt{s} \right)} \\ - \frac{ \frac{r_j r_i}{r_{0_j} r_{0_{i\vert j}}} \exp\left(-\frac{\left( r_{0_j}+r_{0_{i\vert j}} \right)-\left( r_j+r_i \right)}{\sqrt{D}} \sqrt{s} \right) }{ 1-  \frac{r_i r_j}{r_{0_{i\vert j}} r_{0_{j\vert i}}} \exp\left(-\frac{\left( r_{0_{i\vert j}} + r_{0_{j\vert i}} \right)- \left( r_i+r_j \right)}{\sqrt{D}} \sqrt{s} \right)}.
\end{multline}
In order to take the inverse Laplace transform of (\ref{eq:simo_laplace_expanded}), the left-hand part of the fraction needs to be expanded as
\begin{align*}
&\frac{r_i}{r_{0_i}} \exp\left(- \frac{r_{0_i}-r_i}{\sqrt{D}} \sqrt{s} \right) \times \numberthis \label{eq:simo_rec_laplace}\\
&\sum_{k=0}^{\infty} \Bigg[\frac{r_i r_j}{r_{0_{i \vert j}} r_{0_{j \vert i}}} \exp\left(-\frac{(r_{0_{i \vert j}} + r_{0_{j \vert i}})-(r_i+r_j)}{\sqrt{D}} \sqrt{s} \right) \Bigg]^k \\
= &\sum_{k=0}^{\infty} \frac{r_i}{r_{0_i}} \left( \frac{r_i r_j}{r_{0_{i \vert j}} r_{0_{j \vert i}}} \right)^k \times \\ 
&\exp\left( - \frac{\left( r_{0_i}-r_i \right) + k\left( (r_{0_{i \vert j}} + r_{0_{j \vert i}})-(r_i+r_j ) \right)}{\sqrt{D}} \sqrt{s} \right), 
\end{align*}
provided the region of convergence (ROC) is satisfied (ROC is derived in Appendix). Similarly, right-hand part of the fraction is expanded as
\begin{align*}
\sum_{k=0}^{\infty} \frac{r_j r_i}{r_{0_j} r_{0_{i \vert j}}} &\left( \frac{r_i r_j}{r_{0_{i \vert j}} r_{0_{j \vert i}}} \right)^k \exp\Bigg( - \frac{(r_{0_j}+r_{0_{i \vert j}})-(r_i+r_j) }{\sqrt{D}} \\
- &\frac{k\left( (r_{0_{i \vert j}} + r_{0_{j \vert i}})-(r_i+r_j ) \right)}{\sqrt{D}} \sqrt{s} \Bigg). \numberthis
\end{align*}
To further simplify derivations, the expression is converted to
\begin{equation}
P_i^{\text{SIMO}}(s) = \sum_{k=0}^{\infty} A_{ik} \exp\left(-B_{ik}\sqrt{s} \right) - \sum_{k=0}^{\infty} A_{jk} \exp\left(-B_{jk}\sqrt{s} \right),
\end{equation}
where
\begin{align*}
A_{ik} &= \frac{r_i}{r_{0_i}} \left( \frac{r_i r_j}{r_{0_{i \vert j}} r_{0_{j \vert i}}} \right)^k,\ \quad\ A_{jk} = \frac{r_j r_i}{r_{0_j} r_{0_{i \vert j}}} \left( \frac{r_i r_j}{r_{0_{i \vert j}} r_{0_{j \vert i}}} \right)^k,\\
B_{ik} &= \frac{\left( r_{0_i}-r_i \right) + k\left( (r_{0_{j \vert i}} + r_{0_{i \vert j}})-(r_i+r_j ) \right)}{\sqrt{D}},\\
B_{jk} &= \frac{(r_{0_i}+r_{0_{i \vert j}}-r_i-r_j) + k\left( r_{0_{j \vert i}} + r_{0_{i \vert j}}-r_i-r_j \right)}{\sqrt{D}}.
\end{align*}
Due to the linearity property of the Laplace transform, time-domain solution can be expressed as
\begin{multline}
\label{eq:laplace_linearity}
p_{\text{hit}}^{\text{SIMO}}(\text{Rx}_i,t) = \sum_{k=0}^{\infty} \mathcal{L}^{-1}\left\{A_{ik} \exp\left(-B_{ik}\sqrt{s} \right)\right\}\\
- \sum_{k=0}^{\infty} \mathcal{L}^{-1}\left\{A_{jk} \exp\left(-B_{jk}\sqrt{s} \right)\right\}.
\end{multline}
Taking the inverse Laplace transform of each term gives
\begin{equation}
\label{eq:inverse_laplace_single}
\mathcal{L}^{-1}\left\{ A_{ik} \exp\left(-B_{ik}\sqrt{s} \right) \right\} = \frac{A_{ik}}{\sqrt{4\pi t}}\frac{B_{ik}}{t} \exp\left(-\frac{B_{ik}^2}{4t}\right).
\end{equation}
Then, the closed-form solution for molecular SIMO absorption rate of molecules absorbed by Rx$_i$ is obtained as
\begin{multline}
p_{\text{hit}}^{\text{SIMO}}(\text{Rx}_i,t) = \sum_{k=0}^{\infty} \frac{A_{ik}}{\sqrt{4\pi t}}\frac{B_{ik}}{t} \exp\left(-\frac{B_{ik}^2}{4t}\right) \\
- \sum_{k=0}^{\infty} \frac{A_{jk}}{\sqrt{4\pi t}}\frac{B_{jk}}{t} \exp\left(-\frac{B_{jk}^2}{4t}\right).
\label{eq:simo_recursive_clsdform}
\end{multline}
Theoretical closed-form solution consists of the sum of infinite terms. However, since the coefficients $A_{ik}$ and $A_{jk}$ are less than $1$, and the exponents $B_{ik}^2$ and $B_{jk}^2$ grow rapidly which enable the exponential to converge to $0$ rapidly, the expression converges after several terms. To find the absorption probability of the molecules until time $t$, integrating (\ref{eq:simo_recursive_clsdform}) gives
\begin{multline*}
P_{\text{hit}}^{\text{SIMO}}(\text{Rx}_i,t) = \int_0^t p_{\text{hit}}^{\text{SIMO}}(\text{Rx}_i,\tau) d\tau,\\
= \int_0^t \Bigg[ \sum_{k=0}^{\infty} \frac{A_{ik}}{\sqrt{4\pi \tau}}\frac{B_{ik}}{\tau} \exp\left(-\frac{B_{ik}^2}{4\tau}\right) \\
-\sum_{k=0}^{\infty} \frac{A_{jk}}{\sqrt{4\pi \tau}}\frac{B_{jk}}{\tau} \exp\left(-\frac{B_{jk}^2}{4\tau}\right) \Bigg] d\tau,\\
= \sum_{k=0}^{\infty} A_{ik} \erfc\left(\frac{B_{ik}}{\sqrt{4t}} \right) - \sum_{k=0}^{\infty} A_{jk} \erfc\left(\frac{B_{jk}}{\sqrt{4t}} \right),
\end{multline*}
\begin{align*}
= &\sum_{k=0}^{\infty}  \frac{r_i}{r_{0_i}} \left( \frac{r_i r_j}{r_{0_{i \vert j}} r_{0_{j \vert i}}} \right)^k \times \numberthis \label{eq:simo_recursive_cdf}\\
&\erfc\left(\frac{\left( r_{0_i}-r_i \right) + k\left( (r_{0_{j \vert i}} + r_{0_{i \vert j}})-(r_i+r_j ) \right)}{\sqrt{4Dt}} \right)\\
- &\sum_{k=0}^{\infty} \frac{r_j r_i}{r_{0_j} r_{0_{i \vert j}}} \left( \frac{r_i r_j}{r_{0_{i \vert j}} r_{0_{j \vert i}}} \right)^k \erfc\Bigg( \frac{\left( (r_{0_i}+r_{0_{i \vert j}})-(r_i+r_j) \right)}{\sqrt{4Dt}} \\
+&\frac{k\left( (r_{0_{j \vert i}} + r_{0_{i \vert j}})-(r_i+r_j ) \right)}{\sqrt{4Dt}} \Bigg) 
\end{align*}
With the expression in (\ref{eq:simo_recursive_cdf}), the absorption probability of a single molecule until time $t$ is derived. The absorption probability function is a characteristic feature for each receiver in every communication topology.

\subsection{Simplified Model}
\label{subsection:ApproxModel}

Although the recursive model provides a detailed analytical expression for the absorption rates of the molecules towards both receivers, it may become computationally infeasible due to increasing number of terms taken into account. The computation becomes an infinite sum of terms, and the computational complexity increases linearly with the required precision. To simplify this higher-complexity approach, the recursive integral representing the absorption rate of the total of the molecules in (\ref{eq:total_hit}) is \textit{expanded} by one degree by substituting the left-hand side integrand with (\ref{eq:simo_integral}), as
\begin{multline}
\label{eq:simo_general_in_simplified}
p_{\text{hit}}(\text{Rx}_i,t;\ \text{Rx}_j) = \int_{0}^{t} p_{\text{hit}}^{\text{SIMO}}(\text{Rx}_j, \tau) \times \\
p_{\text{hit}}^{\text{SISO}}(\text{Rx}_i,t-\tau \vert r_{0_{i \vert j}},r_i) d\tau,
\end{multline}
\begin{align*}
\label{eq:simo_one_deg_in_simplified}
= &\int_{0}^{t} \Bigg[ p_{\text{hit}}^{\text{SISO}}(\text{Rx}_j,\tau) -
\int_{0}^{\tau} p_{\text{hit}}^{\text{SIMO}}(\text{Rx}_i, \zeta) \times \numberthis\\
&p_{\text{hit}}^{\text{SISO}}(\text{Rx}_j,\tau-\zeta \vert r_{0_{j \vert i}},r_j) d\zeta \Bigg] p_{\text{hit}}^{\text{SISO}}(\text{Rx}_i,t-\tau \vert r_{0_{i \vert j}},r_i) d\tau,
\end{align*}
\begin{multline}
\label{eq:one-deg-expansion}
= \int_{0}^{t} p_{\text{hit}}^{\text{SISO}}(\text{Rx}_j,\tau) p_{\text{hit}}^{\text{SISO}}(\text{Rx}_i,t-\tau \vert r_{0_{i \vert j}},r_i) d\tau - \\
\int_{0}^{t} \int_{0}^{\tau} p_{\text{hit}}^{\text{SIMO}}(\text{Rx}_i, \zeta) p_{\text{hit}}^{\text{SISO}}(\text{Rx}_j,\tau-\zeta \vert r_{0_{j \vert i}},r_j) \times \\
p_{\text{hit}}^{\text{SISO}}(\text{Rx}_i,t-\tau \vert r_{0_{i \vert j}},r_i) d\zeta d\tau.
\end{multline}
The right-hand side double integral corresponds to the effect of the Rx$_i$ on Rx$_j$. With this expansion, the recursive integral is now expanded into \textit{second-degree} terms, which correspond to the multiplication of two absorption rate terms. The double integral in (\ref{eq:one-deg-expansion}) can be further expanded into higher-degree terms to emphasize the effect of reciprocal and recursive effect among receivers, causing even more absorption rate terms multiplying whose result converges to $0$ rapidly. Notice that the conditional-absorption rate terms are accumulated within the expanding integral.

The multiplication of multiple conditional-absorption rate terms tend to converge to $0$, since these terms take values much smaller than $1$. In this case, the double integral that contains the multiplication of multiple conditional-absorption rate terms can be assumed to be $0$, and the latter double integral in (\ref{eq:one-deg-expansion}) can be neglected to arrive at
\begin{multline}
\hat{p}_{\text{hit}}(\text{Rx}_i,t;\ \text{Rx}_j) = \int_{0}^{t} p_{\text{hit}}^{\text{SISO}}(\text{Rx}_j,\tau) \times\\
p_{\text{hit}}^{\text{SISO}}(\text{Rx}_i,t-\tau \vert r_{0_{i \vert j}},r_i) d\tau,
\end{multline}
\begin{multline}
\hat{p}_{\text{hit}}^{\text{SIMO}}(\text{Rx}_i,t) = p_{\text{hit}}^{\text{SISO}}(\text{Rx}_i,t) - \\
\int_{0}^{t} p_{\text{hit}}^{\text{SISO}}(\text{Rx}_j,\tau) p_{\text{hit}}^{\text{SISO}}(\text{Rx}_i,t-\tau \vert r_{0_{i \vert j}},r_i) d\tau.
\end{multline}
As the terms coming from further expansion go to $0$, the simplified expression lies within a close range of the recursive expression. Therefore, molecular SIMO the absorption rate of the molecules towards receivers can be approximated by the simplified model.

The closed-form solution {of the simplified molecular SIMO absorption rate expressions} is quite straightforward compared to its recursive counterpart. It can be {derived} via both convolution integral and Laplace transform. Transforming terms into Laplace domain gives
\begin{equation}
\label{eq:simplified_laplace}
\hat{P}_i^{\text{SIMO}}(s) = P_i^{\text{SISO}}(s) - P_j^{\text{SISO}}(s)P_{i\vert j}^{\text{SISO}}(s).
\end{equation}
Substituting relevant expressions into (\ref{eq:simplified_laplace}) gives
\begin{multline}
\hat{P}_i^{\text{SIMO}}(s) = \frac{r_i}{r_{0_i}} \exp\left(- \frac{r_{0_i}-r_i}{\sqrt{D}} \sqrt{s} \right) \\
- \frac{r_j r_i}{r_{0_j} r_{0_{i \vert j}}} \exp\left(- \frac{(r_{0_j}+r_{0_{i \vert j}})-(r_j+r_i)}{\sqrt{D}} \sqrt{s} \right).
\end{multline}
Inverse-transforming each term into time domain gives
\begin{multline}
\hat{p}_{\text{hit}}^{\text{SIMO}}(\text{Rx}_i,t) = \frac{r_i}{r_{0_i}}\frac{1}{\sqrt{4\pi D t}} \frac{r_{0_i}-r_i}{t}\exp\left(-\frac{(r_{0_i}-r_i)^2}{4Dt}\right)\\
-\frac{r_j r_i}{r_{0_j} r_{0_{i \vert j}}} \frac{1}{\sqrt{4\pi D t}} \frac{(r_{0_j}+r_{0_{i \vert j}})-(r_j+r_i)}{t} \times \\
\exp\left(-\frac{(r_{0_j}+r_{0_{i \vert j}})-(r_j+r_i)^2}{4Dt}\right).
\label{eq:approx_closedform}
\end{multline}
With this expression, a simplification of the recursive model is derived by neglecting the absorption rate terms of recursively affected molecules, which converge to $0$. To find the absorption probability of the molecules until time $t$, integrating (\ref{eq:approx_closedform}) gives
\begin{align*}
\hat{P}_{\text{hit}}^{\text{SIMO}}(\text{Rx}_i,t) = &\int_0^t \hat{p}_{\text{hit}}^{\text{SIMO}}(\text{Rx}_i,\tau) d\tau, \numberthis \label{eq:simo_simplified_cdf}\\
= &\frac{r_i}{r_{0_i}}\erfc\left(\frac{r_{0_i}-r_i}{\sqrt{4Dt}} \right) \\
- &\frac{r_j r_i}{r_{0_j} r_{0_{i \vert j}}} \erfc\left(-\frac{(r_{0_i}+r_{0_{i \vert j}})-(r_i+r_j)}{\sqrt{4Dt}}\right), %
\end{align*}
which is a more computationally feasible closed form solution for the absorption probability of molecules of receivers of a molecular SIMO system.

\subsection{Extension for More Than Two Receivers}
\label{subsection:Extension}
The proposed models can be extended for more than two receivers. In this case, the molecules {absorbed by} all other receivers need to be taken into account. Similar to the $2$-Rx case, the approach is based on the independence of Brownian motion steps. The molecules {absorbed by} other receivers are again assumed to be released from the virtual release points of each receiver as if they were not absorbed, at all. The {joint-absorption rates} of the molecules {absorbed by} other receivers are subtracted from the molecular SISO {absorption rate of molecules towards} the intended {receiver}. To simplify the expression, each molecule is assumed to {be absorbed by} only one receiver before it is {considered to be} virtually released to the intended {receiver.} 

Without loss of generality, the molecular SIMO response can be expressed for Rx$_1$ in a $n$-Rx molecular SIMO topology. The joint absorption probability of molecules that first hit Rx$_j$ at time $\tau$, then hit Rx$_1$ at time $t>\tau$ is
\begin{equation}
\label{eq:multi_rx_first_eqn}
p_{\text{hit}}(\text{Rx}_1,t;\text{Rx}_j,\tau) = p_{\text{hit}}^{\text{SIMO}}(\text{Rx}_j,\tau) p_{\text{hit}}(\text{Rx}_1,t \vert \text{Rx}_j, \tau).
\end{equation}
Integrating over all $\tau < t$ gives the impact of Rx$_j$ on Rx$_1$ as
\begin{equation}
p_{\text{hit}}(\text{Rx}_1,t;\text{Rx}_j) =  \int_{0}^{t} p_{\text{hit}}^{\text{SIMO}}(\text{Rx}_j,\tau) p_{\text{hit}}(\text{Rx}_1,t \vert \text{Rx}_j, \tau)d\tau.
\end{equation}
The impacts of each receiver on Rx$_1$ needs to be subtracted from the molecular SISO response of Rx$_1$, therefore the molecular SIMO response of Rx$_1$ is
\begin{multline}
\label{eq:multiple_integral_1}
p_{\text{hit}}^{\text{SIMO}}(\text{Rx}_1,t) = p_{\text{hit}}^{\text{SISO}}(\text{Rx}_1,t) \\
- \sum_{j=2}^{n} \int_{0}^{t} p_{\text{hit}}^{\text{SIMO}}(\text{Rx}_j,\tau) p_{\text{hit}}(\text{Rx}_1,t \vert \text{Rx}_j, \tau)d\tau.
\end{multline}
The general recursive molecular SIMO response expression is thus
\begin{multline}
\label{eq:multiple_integral}
p_{\text{hit}}^{\text{SIMO}}(\text{Rx}_i,t) = p_{\text{hit}}^{\text{SISO}}(\text{Rx}_i,t) \\
- \sum_{\substack{j=1 \\ j\neq i}}^{n} \int_{0}^{t} p_{\text{hit}}^{\text{SIMO}}(\text{Rx}_j,\tau) p_{\text{hit}}(\text{Rx}_i,t \vert \text{Rx}_j, \tau)d\tau.
\end{multline}
The expression can be transformed into the Laplace domain in a similar fashion, as
\begin{equation}
P_i^{\text{SIMO}}(s) = P_i^{\text{SISO}}(s) - \sum_{\substack{j=1 \\ j\neq i}}^{n} P_j^{\text{SIMO}}(s) P_{i\vert j}^{\text{SISO}}(s).
\end{equation}
Gathering the molecular SIMO terms in one side gives
\begin{equation}
P_i^{\text{SIMO}}(s) + \sum_{\substack{j=1 \\ j\neq i}}^{n} P_j^{\text{SIMO}}(s) P_{i\vert j}^{\text{SISO}}(s) = P_i^{\text{SISO}}(s),
\end{equation}
which can be interpreted as a linear system. The linear system can be shaped into matrix notation as
\begin{multline}
\begin{bmatrix}
1 & P_{{1\vert 2}}^{\text{SISO}}(s) & \dots & P_{{1\vert n_\text{Rx}}}^{\text{SISO}}(s)\\
P_{{2\vert 1}}^{\text{SISO}}(s) & 1 & \dots & P_{{2\vert n_\text{Rx}}}^{\text{SISO}}(s)\\
\vdots & \vdots & \ddots & \vdots\\
P_{{n_\text{Rx}\vert 1}}^{\text{SISO}}(s) & P_{{n_\text{Rx}\vert 2}}^{\text{SISO}}(s) & \dots & 1
\end{bmatrix} \times\\
\begin{bmatrix}
P_{1}^{\text{SIMO}}(s)\\
P_{2}^{\text{SIMO}}(s)\\
\vdots\\
P_{{n_\text{Rx}}}^{\text{SIMO}}(s)\\
\end{bmatrix} = \begin{bmatrix}
P_{1}^{\text{SISO}}(s)\\
P_{2}^{\text{SISO}}(s)\\
\vdots\\
P_{{n_\text{Rx}}}^{\text{SISO}}(s)
\end{bmatrix}.
\end{multline}
Although it is not a trivial task, solving the linear system would yield the Laplace transforms of the absorption rate of molecules for each receivers. The recursive integral in \eqref{eq:multiple_integral} can be approximated, similar to the $2$-Rx case, by leaving only the molecular SISO response in the integrand as the most dominant term, which gives the approximated model as
\begin{multline}
p_{\text{hit}}^{\text{SIMO}}(\text{Rx}_i,t) \approx p_{\text{hit}}^{\text{SISO}}(\text{Rx}_i,t) \\
- \sum_{\substack{j=1 \\ j\neq i}}^{n} \int_{0}^{t} p_{\text{hit}}^{\text{SISO}}(\text{Rx}_j,\tau) p_{\text{hit}}(\text{Rx}_i,t \vert \text{Rx}_j, \tau)d\tau.
\label{eq:multiapprox_integral}
\end{multline}
The expression again can be transformed into the Laplace domain as
\begin{equation}
P_i^{\text{SIMO}}(s) = P_i^{\text{SISO}}(s) - \sum_{\substack{j=1 \\ j\neq i}}^{n} P_j^{\text{SISO}}(s) P_{i\vert j}^{\text{SISO}}(s),
\end{equation}
which can be shaped in matrix notation as
\begin{multline}
\begin{bmatrix}
1 & -P_{{1\vert 2}}^{\text{SISO}}(s) & \dots & -P_{{1\vert n_\text{Rx}}}^{\text{SISO}}(s)\\
-P_{{2\vert 1}}^{\text{SISO}}(s) & 1 & \dots & -P_{{2\vert n_\text{Rx}}}^{\text{SISO}}(s)\\
\vdots & \vdots & \ddots & \vdots\\
-P_{{n_\text{Rx}\vert 1}}^{\text{SISO}}(s) & -P_{{n_\text{Rx}\vert 2}}^{\text{SISO}}(s) & \dots & 1
\end{bmatrix} \times \\ \begin{bmatrix}
P_{1}^{\text{SISO}}(s)\\
P_{2}^{\text{SISO}}(s)\\
\vdots\\
P_{{n_\text{Rx}}}^{\text{SISO}}(s)
\end{bmatrix} = \begin{bmatrix}
\hat{P}_{1}^{\text{SIMO}}(s)\\
\hat{P}_{2}^{\text{SIMO}}(s)\\
\vdots\\
\hat{P}_{{n_\text{Rx}}}^{\text{SIMO}}(s)\\
\end{bmatrix},
\end{multline}
and the Laplace transforms of the closed-form solutions of the absorption rates of molecules for each receiver can be approximated by simply solving the linear system formulated by the molecular SISO analytics based on the spatial parameters.

\section{Performance Evaluation}
\label{section:Performance}

\subsection{Angular Analysis}
\label{subsection:angular_analysis}

Several topology parameters become of importance regarding the definition of a 2-Rx molecular SIMO topology. Three non-collinear points always define a plane in 3-D space. The transmitter and the centers of the {two} receivers define a plane in space, and the topology parameters are defined on the aforementioned plane. For a Tx-centralized perspective, these parameters are $r_1, r_2,r_{0_1}, r_{0_2}$, and the separation angle $\phi$, which defines the angular separation between centers of Rxs with respect to Tx, presented in Fig.~\ref{fig:simo_model}. Varying parameters define a range of different-responding topologies, so it is essential to examine the estimation performance of the model for several topologies with different parameter sets. The analyses for this study are conducted on three different scenarios that are designed based on the relative sizes of the receivers. Channel response of different topology examples are obtained by computer-based simulations, {particle-based diffusion simulations}, and evaluated by the comprehensive recursive model. {The simulations are conducted with number of molecules of $N = \SI{e6}{}$, time step size of $\Delta t = \SI{e-4}{\second}$ with $10$ iterations. The channel responses are acquired by averaging all iterations, and normalizing with respect to $N$.} Performance evaluations and comparisons are conducted between these simulated channel responses and the channel responses obtained by the comprehensive recursive model.

For small separation angles, the closer Rx blocks the sight of Tx towards the farther Rx. This causes that the farther Rx is not fully covered due to the physical presence of the closer Rx. This blockage is called the shadowing phenomenon throughout this paper. The {absorption probability of molecules by} the farther Rx is affected due to the shadowing, especially for the angles closer to $0$.

Two significant angles are examined throughout this phenomenon, namely the half-eclipse angle $\varphi_1$ and the no-eclipse angle $\varphi_2$. The half-eclipse angle corresponds to the angle at which the center of the farther Rx lies on the {edge of the tangent cone of the closer Rx}, causing roughly one half of the farther Rx to be shadowed by the closer Rx. It is calculated by $\varphi_1 = \arcsin {\frac{r_1}{r_{0_1}}}\ \SI{}{\radian}$. The no-eclipse angle corresponds to the angle at which {the edges of the tangent cones of} both Rxs coincide and there is no shadowing, {which is calculated by $\varphi_2 = \arcsin {\frac{r_1}{r_{0_1}}}+\arcsin {\frac{r_2}{r_{0_2}}}\ \SI{}{\radian}$.}

\subsubsection{Scenario 1 -- Smaller Rx to be closer to Tx}
\label{subsubsec:scenario1}
\begin{table}[ht]
\normalsize
\centering
\caption{\normalsize Topology Parameters for Scenario 1}
\begin{tabular}{c||c c c c}
\hline\hline
Parameter & $r_1$ & $r_2$ & $r_{0_1}$ & $r_{0_2}$ \\ \hline
Value & \SI{2}{\micro\meter} & \SI{5}{\micro\meter} & \SI{6}{\micro\meter} & \SI{16}{\micro\meter}\\ \hline\hline
\end{tabular}
\label{table:sc1}
\end{table}

This configuration offers a topology where the Rx closer to Tx is smaller, and cannot completely shadow the farther Rx when the separation angle $\phi$ is close or equal to $0$. This allows the farther Rx to still receive molecules even if it is eclipsed by the closer Rx. The closest distance between receivers is set to $\SI{3}{\micro\meter}$, in case they are perfectly aligned, namely $\phi = 0$. This allows the release point from the farther Rx to be {relatively distant} to the center of the closer Rx.

\begin{figure}[ht]
\centering
\begin{tikzpicture}
\begin{axis}[ width=\linewidth, height=.5\linewidth,
    xlabel={Separation angle $(\deg)$},
    ylabel={RMS error}, ymin=0, grid,
    xtick={0, 60, 90, 120, 150, 180},
    extra x ticks={19.47, 37.68},
    extra x tick labels={$\varphi_1$,$\varphi_2$},
    legend style={at={(0.98,0.98)},anchor=north east,
    every axis plot/.append style={very thick},
    nodes={scale=0.4, transform shape}} ]
\addplot[color=blue,mark=o] table[x=angles, y=rms_exact1]{data/r1_2_r2_5_perf.txt};
\addplot[color=red,mark=o] table[x=angles,y=rms_exact2]        {data/r1_2_r2_5_perf.txt};
\addplot[color=blue,mark=square*,fill=blue,only marks]        table[x=angles,y=rms_exact1]      {data/r1_2_r2_5_halfEclipse.txt};
\addplot[color=red,mark=square*,fill=red,only marks]
    table[x=angles, y=rms_exact2]        {data/r1_2_r2_5_halfEclipse.txt};
\addplot[color=blue,mark=triangle*,fill=blue,only marks]
    table[x=angles,y=rms_exact1]        {data/r1_2_r2_5_noEclipse.txt};
\addplot[color=red,mark=triangle*,fill=red,only marks]
    table[x=angles,y=rms_exact2]        {data/r1_2_r2_5_noEclipse.txt};
\draw (19.471,0) -- (19.471,3e-2);
\draw (37.681,0) -- (37.681,3e-2);
\legend{RMS error of Rx$_1$ (Comp. Model), RMS error of Rx$_2$ (Comp. Model), RMS error of Rx$_1$ at half eclipse angle, RMS error of Rx$_2$ at half eclipse angle,RMS error of Rx$_1$ at no eclipse angle,RMS error of Rx$_1$ at no eclipse angle}
\end{axis}
\end{tikzpicture}%
\caption{Plots of RMS error of Rx$_1$ and Rx$_2$ for Scenario 1, with variable angular separation, $\varphi_1 = 19.47\ \SI{}{\deg}$, $\varphi_2 = 37.68\ \SI{}{\deg}$.}
\label{fig:rms_scenario1}
\end{figure}

\begin{figure*}[ht]
\centering
\subcaptionbox{Plots of absorption probability of molecules of Rx$_1$ and Rx$_2$ for Scenario 1 placed with $\varphi_1 = \arcsin {\frac{r_1}{r_{0_1}}}\ \SI{}{\radian}$ angular separation.}%
[.32\textwidth]{
\begin{tikzpicture}
\begin{axis}[ width=\linewidth, height=1.2\linewidth, xlabel={Time $(s)$},
    ylabel={Absorption probability of molecules},
    ymin=0, ymax=0.4, xmin=0, xmax=5, grid,
    every axis plot/.append style={very thick},
    y tick label style={/pgf/number format/.cd,
    fixed, fixed zerofill, precision=2},
    legend pos=north east,
    legend style={nodes={scale=0.4, transform shape}} ]
\addplot[color=blue, postaction={ decoration={markings,
      mark=between positions 0 and 1 step 0.1 with { \fill circle[radius=2pt]; },}, decorate,}, forget plot] table[x=t,y=exact1]
    {data/r1_2_r2_5_r01_6_r02_16_halfEclipse_cdf.txt};
    \addlegendimage{color=blue, postaction={decoration={ markings, mark=between positions .5 and .5 step .5 with { \fill circle[radius=2pt]; },}, decorate,} }
\addplot[color=red, postaction={ decoration={markings,
      mark=between positions 0 and 1 step 0.1 with { \fill circle[radius=2pt]; },}, decorate,},forget plot] table[x=t,y=sim1]
    {data/r1_2_r2_5_r01_6_r02_16_halfEclipse_cdf.txt};
    \addlegendimage{color=red, postaction={decoration={ markings, mark=between positions .5 and .5 step .5 with { \fill circle[radius=2pt]; },}, decorate,} }
\addplot[color=teal, postaction={ decoration={markings,
      mark=between positions 0 and 1 step 0.1 with { \fill circle[radius=2pt]; },}, decorate,},forget plot]  table[x=t,y=exact2]
    {data/r1_2_r2_5_r01_6_r02_16_halfEclipse_cdf.txt};
    \addlegendimage{color=teal, postaction={decoration={ markings, mark=between positions .5 and .5 step .5 with { \fill circle[radius=2pt]; },}, decorate,} }
\addplot[color=orange, postaction={ decoration={markings,
      mark=between positions 0 and 1 step 0.1 with { \fill circle[radius=2pt]; },}, decorate,},forget plot] table[x=t,y=sim2]
    {data/r1_2_r2_5_r01_6_r02_16_halfEclipse_cdf.txt};
    \addlegendimage{color=orange, postaction={decoration={ markings, mark=between positions .5 and .5 step .5 with { \fill circle[radius=2pt]; },}, decorate,} }
\addplot[color=blue, dashed] table[x=time,y=sc1_r1]
    {data/all_scs_siso_cdf.txt};
\addplot[color=teal, dashed] table[x=time,y=sc1_r2]
    {data/all_scs_siso_cdf.txt};
\legend{Molecular SIMO response of Rx$_1$ (Comp. Model), Molecular SIMO response of Rx$_1$ (Simulation), Molecular SIMO response of Rx$_2$ (Comp. Model), Molecular SIMO response of Rx$_2$ (Simulation)}   
\end{axis}
\end{tikzpicture} \label{fig:cdfs_scenario1_half}
} 
\subcaptionbox{Plots of absorption probability of molecules of Rx$_1$ and Rx$_2$ for Scenario 1 placed with $\frac{\pi}{2}\ \SI{}{\radian}$ angular separation.}%
[.32\textwidth]{
\begin{tikzpicture}
\begin{axis}[ width=\linewidth, height=1.2\linewidth, xlabel={Time $(s)$},
    ylabel={Absorption probability of molecules},
    ymin=0, ymax=0.4, xmin=0, xmax=5, grid,
    every axis plot/.append style={very thick},
    y tick label style={/pgf/number format/.cd,
    fixed, fixed zerofill, precision=2 },
    legend pos=north east,
    legend style={nodes={scale=0.4, transform shape}} ]
\addplot[color=blue, postaction={ decoration={markings,
      mark=between positions 0 and 1 step 0.1
           with { \fill circle[radius=2pt]; },},
    decorate,},forget plot] table[x=t,y=exact1]
    {data/r1_2_r2_5_r01_6_r02_16_90deg_cdf.txt};
    \addlegendimage{color=blue, postaction={decoration={ markings, mark=between positions .5 and .5 step .5 with { \fill circle[radius=2pt]; },}, decorate,} }
\addplot[color=red, postaction={ decoration={markings,
      mark=between positions 0 and 1 step 0.1
           with { \fill circle[radius=2pt]; },},
    decorate,},forget plot] table[x=t,y=sim1]
    {data/r1_2_r2_5_r01_6_r02_16_90deg_cdf.txt};
    \addlegendimage{color=red, postaction={decoration={ markings, mark=between positions .5 and .5 step .5 with { \fill circle[radius=2pt]; },}, decorate,} }
\addplot[color=teal, postaction={ decoration={markings,
      mark=between positions 0 and 1 step 0.1
           with { \fill circle[radius=2pt]; },},
    decorate,},forget plot] table[x=t,y=exact2]
    {data/r1_2_r2_5_r01_6_r02_16_90deg_cdf.txt};
    \addlegendimage{color=teal, postaction={decoration={ markings, mark=between positions .5 and .5 step .5 with { \fill circle[radius=2pt]; },}, decorate,} }
\addplot[color=orange, postaction={ decoration={markings,
      mark=between positions 0 and 1 step 0.1
           with { \fill circle[radius=2pt]; },},
    decorate,},forget plot] table[x=t,y=sim2]
    {data/r1_2_r2_5_r01_6_r02_16_90deg_cdf.txt};
    \addlegendimage{color=orange, postaction={decoration={ markings, mark=between positions .5 and .5 step .5 with { \fill circle[radius=2pt]; },}, decorate,} }
\addplot[color=blue, dashed]
    table[x=time,y=sc1_r1]
    {data/all_scs_siso_cdf.txt};
\addplot[color=teal, dashed]
    table[x=time,y=sc1_r2]
    {data/all_scs_siso_cdf.txt};
\legend{Molecular SIMO response of Rx$_1$ (Comp. Model), Molecular SIMO response of Rx$_1$ (Simulation), Molecular SIMO response of Rx$_2$ (Comp. Model), Molecular SIMO response of Rx$_2$ (Simulation)}  
\end{axis}
\end{tikzpicture} \label{fig:cdfs_scenario1_pi_2}
} 
\subcaptionbox{Plots of absorption probability of molecules of Rx$_1$ and Rx$_2$ for Scenario 1 placed with $\pi\ \SI{}{\radian}$ angular separation.}%
[.32\textwidth]{
\begin{tikzpicture}
\begin{axis}[ width=\linewidth, height=1.2\linewidth, xlabel={Time $(s)$},
    ylabel={Absorption probability of molecules},
    ymin=0, ymax=0.4, xmin=0, xmax=5, grid,
    every axis plot/.append style={very thick},
    y tick label style={ /pgf/number format/.cd,
    fixed, fixed zerofill, precision=2 },
    legend pos=north east,
    legend style={nodes={scale=0.4, transform shape}}]
\addplot[color=blue, postaction={ decoration={markings,
      mark=between positions 0 and 1 step 0.1
           with { \fill circle[radius=2pt]; },},
    decorate,},forget plot] table[x=t,y=exact1]
    {data/r1_2_r2_5_r01_6_r02_16_180deg_cdf.txt};
    \addlegendimage{color=blue, postaction={decoration={ markings, mark=between positions .5 and .5 step .5 with { \fill circle[radius=2pt]; },}, decorate,} }
\addplot[color=red, postaction={ decoration={markings,
      mark=between positions 0 and 1 step 0.1
           with { \fill circle[radius=2pt]; },},
    decorate,},forget plot] table[x=t,y=sim1]
    {data/r1_2_r2_5_r01_6_r02_16_180deg_cdf.txt};
    \addlegendimage{color=red, postaction={decoration={ markings, mark=between positions .5 and .5 step .5 with { \fill circle[radius=2pt]; },}, decorate,} }
\addplot[color=teal, postaction={ decoration={markings,
      mark=between positions 0 and 1 step 0.1
           with { \fill circle[radius=2pt]; },},
    decorate,},forget plot] table[x=t,y=exact2]
    {data/r1_2_r2_5_r01_6_r02_16_180deg_cdf.txt};
    \addlegendimage{color=teal, postaction={decoration={ markings, mark=between positions .5 and .5 step .5 with { \fill circle[radius=2pt]; },}, decorate,} }
\addplot[color=orange, postaction={ decoration={markings,
      mark=between positions 0 and 1 step 0.1
           with { \fill circle[radius=2pt]; },},
    decorate,},forget plot] table[x=t,y=sim2]
    {data/r1_2_r2_5_r01_6_r02_16_180deg_cdf.txt};
    \addlegendimage{color=orange, postaction={decoration={ markings, mark=between positions .5 and .5 step .5 with { \fill circle[radius=2pt]; },}, decorate,} }
\addplot[color=blue, dashed]
    table[x=time,y=sc1_r1]
    {data/all_scs_siso_cdf.txt};
\addplot[color=teal, dashed]
    table[x=time,y=sc1_r2]
    {data/all_scs_siso_cdf.txt};
\legend{Molecular SIMO response of Rx$_1$ (Comp. Model), Molecular SIMO response of Rx$_1$ (Simulation), Molecular SIMO response of Rx$_2$ (Comp. Model), Molecular SIMO response of Rx$_2$ (Simulation)} 
\end{axis}
\end{tikzpicture} \label{fig:cdfs_scenario1_pi}
} 
\caption{Plots of absorption probability of molecules of Rx$_1$ and Rx$_2$ for Scenario 1.}
\end{figure*}

\begin{figure}[ht]
\centering
\subcaptionbox{Visualization of the topology for Scenario 1.}%
[.45\linewidth]{
\includegraphics[width=\linewidth]{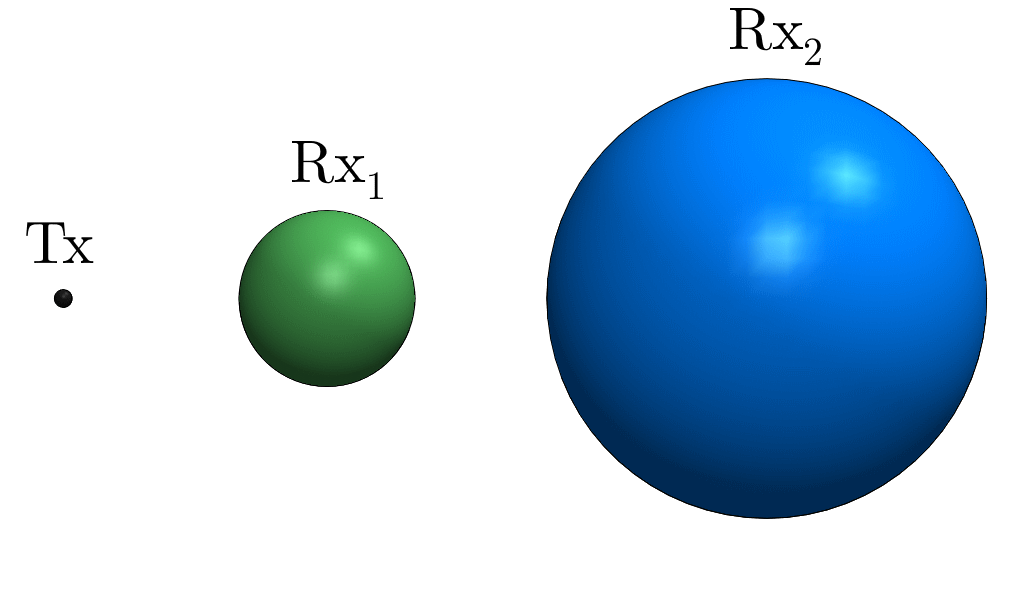}
\label{fig:scenario1_topology}
} 
\hfill
\subcaptionbox{Visualization of the topology for Scenario 2.}%
[.45\linewidth]{
\includegraphics[width=\linewidth]{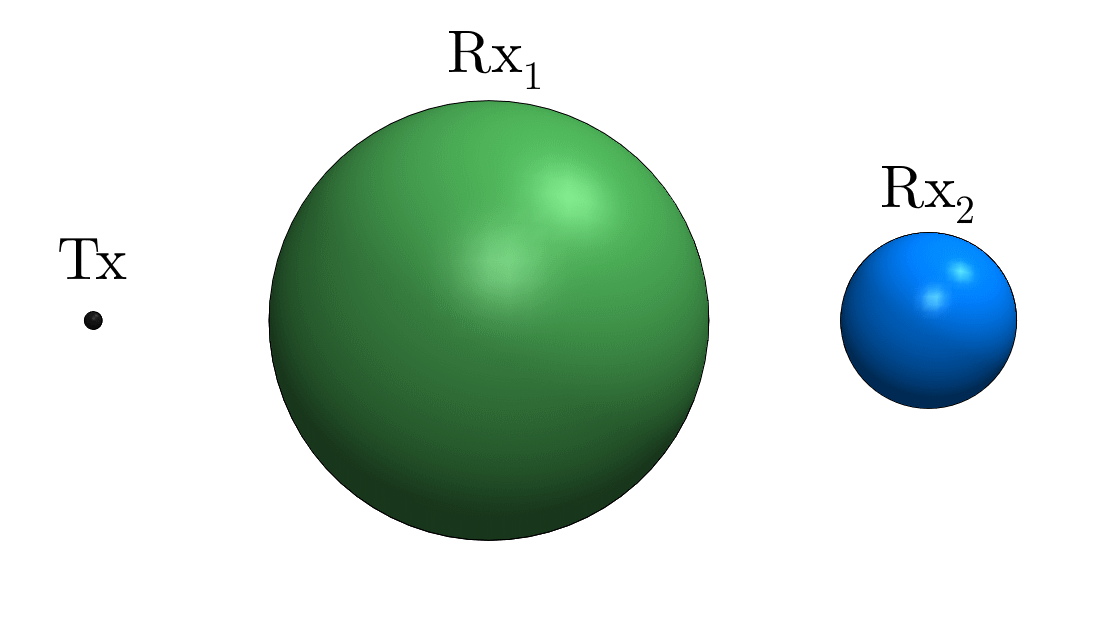}
\label{fig:scenario2_topology}
} 
\caption{Visualizations of the topologies for Scenario 1 and Scenario 2.}
\end{figure}

As the separation angle increases, the shadowing on the farther Rx, namely Rx$_2$, decreases. Also, the release point towards Rx$_2$ from Rx$_1$ becomes more direct, since the virtual release point {of} Rx$_1$ faces Tx. This ensures that as the separation angle increases, root-mean-square (RMS) error between simulation data and recursive model of the absorption probability of molecules absorbed by Rx$_2$ decreases. On the other hand, RMS error on Rx$_1$ increases with increasing separation angle, as shown in Fig.~\ref{fig:rms_scenario1}. This is because of the fact that the virtual release point stands closer to Rx$_1$, and exponential terms of the absorption rate of the stolen molecules have a significant effect on Rx$_1$. With the reducing eclipse, as can be seen until the no-eclipse angle, the RMS error constantly decreases, which shows the performance increase when eclipsing reduces. After the no-eclipse angle, the impact on Rxs gradually disappears, and RMS error saturates. As expected, RMS error on Rx$_2$ constantly decreases as it gets away from the impact of Rx$_1$.

{As expected, the exact absorption point of a molecule on an Rx is random, therefore the virtual release point arises from a generalization of angular absorption distributions.} Since the larger Rx, Rx$_2$ in the topology, is placed further, the distance between the virtual release point and the center of Rx$_2$ is expectedly large, which weakens the exponential terms in the absorption rate of the molecules stolen by Rx$_1$. Therefore the subtracted amount from the absorption rate of Rx$_1$ would be lower. This is the potential reason for the deviation in the channel response of the Rx$_2$.


\subsubsection{Scenario 2 -- Larger Rx to be closer to Tx}
\label{subsubsec:scenario2}
\begin{table}[ht]
\normalsize
\centering
\caption{\normalsize Topology Parameters for Scenario 2}
\begin{tabular}{c||c c c c}
\hline\hline
Parameter & $r_1$ & $r_2$ & $r_{0_1}$ & $r_{0_2}$ \\ \hline
Value & \SI{5}{\micro\meter} & \SI{2}{\micro\meter} & \SI{9}{\micro\meter} & \SI{19}{\micro\meter}\\ \hline\hline
\end{tabular}
\label{table:sc2}
\end{table}

This configuration offers a topology where the Rx closer to Tx is larger, and can completely shadow the farther Rx when the separation angle $\phi$ is close or equal to $0$. This prevents the farther Rx to receive molecules significantly in case it is eclipsed by the closer Rx. The closest distance between receivers is set $\SI{3}{\micro\meter}$, in case they are perfectly aligned, namely $\phi = 0$. This again allows the release point from the farther Rx to be {relatively distant} to the center of the closer Rx.

\begin{figure}[ht]
\centering
\begin{tikzpicture}
\begin{axis}[width=\linewidth, height=.5\linewidth,
    xlabel={Separation angle $(\deg)$}, ylabel={RMS error},
    ymin=0, grid, every axis plot/.append style={very thick},
    xtick={0, 60, 90, 120, 150, 180},
    extra x ticks={32, 42}, extra x tick labels={33.75, 39.79},
    extra x tick labels={$\varphi_1$,$\varphi_2$},
    legend style={at={(0.98,0.4)},anchor=east,
    nodes={scale=0.4, transform shape}}]
\addplot[color=blue,mark=o] table[x=angles,y=rms_exact1] {data/r1_5_r2_2_perf.txt};
\addplot[color=red,mark=o] table[x=angles,y=rms_exact2] {data/r1_5_r2_2_perf.txt};
\addplot[color=blue,mark=square*,fill=blue,only marks]
    table[x=angles,y=rms_exact1]
    {data/r1_5_r2_2_halfEclipse.txt};
\addplot[color=red,mark=square*,fill=red,only marks]
    table[x=angles,y=rms_exact2]
    {data/r1_5_r2_2_halfEclipse.txt};
\addplot[color=blue,mark=triangle*,fill=blue,only marks]
    table[x=angles,y=rms_exact1]
    {data/r1_5_r2_2_noEclipse.txt};
\addplot[color=red,mark=triangle*,fill=red,only marks]
    table[x=angles,y=rms_exact2]
    {data/r1_5_r2_2_noEclipse.txt};
\draw (33.749,0) -- (33.749,3e-2);
\draw (39.791,0) -- (39.791,3e-2);
\legend{RMS error of Rx$_1$ (Comp. Model), RMS error of Rx$_2$ (Comp. Model), RMS error of Rx$_1$ at half eclipse angle, RMS error of Rx$_2$ at half eclipse angle,RMS error of Rx$_1$ at no eclipse angle,RMS error of Rx$_1$ at no eclipse angle} 
\end{axis}
\end{tikzpicture}%
\caption{Plots of RMS error of Rx$_1$ and Rx$_2$ for Scenario 2, with variable angular separation, $\varphi_1 = 33.75\ \SI{}{\deg}$, $\varphi_2 = 39.79\ \SI{}{\deg}$.}
\label{fig:rms_scenario2}
\end{figure}

\begin{figure*}[ht]
\centering
\subcaptionbox{Plots of absorption probability of molecules of Rx$_1$ and Rx$_2$ for Scenario 2 placed with $\varphi_1 = \arcsin {\frac{r_1}{r_{0_1}}}\ \SI{}{\radian}$ angular separation.}%
[.32\textwidth]{
\begin{tikzpicture}
\begin{axis}[width=\linewidth, height=1.2\linewidth,
    xlabel={Time $(s)$},
    ylabel={Absorption probability of molecules},
    ymin=0, ymax=0.55, xmin=0, xmax=5, grid,
    every axis plot/.append style={very thick},
    y tick label style={/pgf/number format/.cd,
    fixed, fixed zerofill, precision=2 },
    legend style={at={(0.98,0.4)},anchor=east,
    nodes={scale=0.4, transform shape} }]
\addplot[color=blue, postaction={decoration={markings,
      mark=between positions 0 and 1 step 0.1 with { \fill circle[radius=2pt]; },}, decorate,},forget plot] table[x=t,y=exact1]     {data/r1_5_r2_2_r01_9_r02_19_halfEclipse_cdf.txt};
      \addlegendimage{color=blue, postaction={decoration={ markings, mark=between positions .5 and .5 step .5 with { \fill circle[radius=2pt]; },}, decorate,} }
\addplot[color=red, postaction={decoration={markings,
      mark=between positions 0 and 1 step 0.1 with { \fill circle[radius=2pt]; },}, decorate,},forget plot] table[x=t,y=sim1]
    {data/r1_5_r2_2_r01_9_r02_19_halfEclipse_cdf.txt};
    \addlegendimage{color=red, postaction={decoration={ markings, mark=between positions .5 and .5 step .5 with { \fill circle[radius=2pt]; },}, decorate,} }
\addplot[color=teal,postaction={decoration={markings,
      mark=between positions 0 and 1 step 0.1 with { \fill circle[radius=2pt]; },}, decorate,},forget plot] table[x=t,y=exact2]
    {data/r1_5_r2_2_r01_9_r02_19_halfEclipse_cdf.txt};
    \addlegendimage{color=teal, postaction={decoration={ markings, mark=between positions .5 and .5 step .5 with { \fill circle[radius=2pt]; },}, decorate,} }
\addplot[color=orange,postaction={decoration={markings,
      mark=between positions 0 and 1 step 0.1 with { \fill circle[radius=2pt]; },}, decorate,},forget plot] table[x=t,y=sim2]
    {data/r1_5_r2_2_r01_9_r02_19_halfEclipse_cdf.txt};
    \addlegendimage{color=orange, postaction={decoration={ markings, mark=between positions .5 and .5 step .5 with { \fill circle[radius=2pt]; },}, decorate,} }
\addplot[color=blue, dashed]
    table[x=time,y=sc2_r1]
    {data/all_scs_siso_cdf.txt};
\addplot[color=teal, dashed]
    table[x=time,y=sc2_r2]
    {data/all_scs_siso_cdf.txt};
\legend{Molecular SIMO response of Rx$_1$ (Comp. Model), Molecular SIMO response of Rx$_1$ (Simulation), Molecular SIMO response of Rx$_2$ (Comp. Model), Molecular SIMO response of Rx$_2$ (Simulation)}    
\end{axis}
\end{tikzpicture} \label{fig_cdfs_scenario2_half}
} 
\subcaptionbox{Plots of absorption probability of molecules of Rx$_1$ and Rx$_2$ for Scenario 2 placed with $\frac{\pi}{2}\ \SI{}{\radian}$ angular separation.}%
[.32\textwidth]{
\begin{tikzpicture}
\begin{axis}[width=\linewidth, height=1.2\linewidth,
    xlabel={Time $(s)$}, ylabel={Absorption probability of molecules}, ymin=0, ymax=0.55, xmin=0, xmax=5, grid,
    every axis plot/.append style={very thick},
    y tick label style={/pgf/number format/.cd,
    fixed, fixed zerofill, precision=2 },
    legend style={at={(0.98,0.4)},anchor=east,
    nodes={scale=0.4, transform shape} }]
\addplot[color=blue,postaction={decoration={markings,
      mark=between positions 0 and 1 step 0.1 with { \fill circle[radius=2pt]; },}, decorate,},forget plot] table[x=t,y=exact1]
    {data/r1_5_r2_2_r01_9_r02_19_90deg_cdf.txt};
    \addlegendimage{color=blue, postaction={decoration={ markings, mark=between positions .5 and .5 step .5 with { \fill circle[radius=2pt]; },}, decorate,} }
\addplot[color=red,postaction={decoration={markings,
      mark=between positions 0 and 1 step 0.1 with { \fill circle[radius=2pt]; },}, decorate,},forget plot] table[x=t,y=sim1]
    {data/r1_5_r2_2_r01_9_r02_19_90deg_cdf.txt};
    \addlegendimage{color=red, postaction={decoration={ markings, mark=between positions .5 and .5 step .5 with { \fill circle[radius=2pt]; },}, decorate,} }
\addplot[color=teal,postaction={decoration={markings,
      mark=between positions 0 and 1 step 0.1 with { \fill circle[radius=2pt]; },}, decorate,},forget plot] table[x=t,y=exact2]
    {data/r1_5_r2_2_r01_9_r02_19_90deg_cdf.txt};
    \addlegendimage{color=teal, postaction={decoration={ markings, mark=between positions .5 and .5 step .5 with { \fill circle[radius=2pt]; },}, decorate,} }
\addplot[color=orange,postaction={decoration={markings,
      mark=between positions 0 and 1 step 0.1 with { \fill circle[radius=2pt]; },}, decorate,},forget plot] table[x=t,y=sim2]
    {data/r1_5_r2_2_r01_9_r02_19_90deg_cdf.txt};
    \addlegendimage{color=orange, postaction={decoration={ markings, mark=between positions .5 and .5 step .5 with { \fill circle[radius=2pt]; },}, decorate,} }
\addplot[color=blue, dashed] table[x=time,y=sc2_r1]
    {data/all_scs_siso_cdf.txt};
\addplot[color=teal, dashed] table[x=time,y=sc2_r2]
    {data/all_scs_siso_cdf.txt};
\legend{Molecular SIMO response of Rx$_1$ (Comp. Model), Molecular SIMO response of Rx$_1$ (Simulation), Molecular SIMO response of Rx$_2$ (Comp. Model), Molecular SIMO response of Rx$_2$ (Simulation)}   
\end{axis}
\end{tikzpicture} \label{fig:cdfs_scenario2_pi_2}
} 
\subcaptionbox{Plots of absorption probability of molecules of Rx$_1$ and Rx$_2$ for Scenario 2 placed with $\pi\ \SI{}{\radian}$ angular separation.}%
[.32\textwidth]{
\begin{tikzpicture}
\begin{axis}[width=\linewidth, height=1.2\linewidth,
    xlabel={Time $(s)$}, ylabel={Absorption probability of molecules}, ymin=0, ymax=0.55, xmin=0, xmax=5, grid,
    every axis plot/.append style={very thick},
    y tick label style={/pgf/number format/.cd,
    fixed, fixed zerofill, precision=2 },
    legend style={at={(0.98,0.4)},anchor=east,
    nodes={scale=0.4, transform shape} }]
\addplot[color=blue,postaction={decoration={markings,
      mark=between positions 0 and 1 step 0.1 with { \fill circle[radius=2pt]; },}, decorate,},forget plot] table[x=t,y=exact1]
    {data/r1_5_r2_2_r01_9_r02_19_180deg_cdf.txt};
    \addlegendimage{color=blue, postaction={decoration={ markings, mark=between positions .5 and .5 step .5 with { \fill circle[radius=2pt]; },}, decorate,} }
\addplot[color=red,postaction={decoration={markings,
      mark=between positions 0 and 1 step 0.1 with { \fill circle[radius=2pt]; },}, decorate,},forget plot] table[x=t,y=sim1]
    {data/r1_5_r2_2_r01_9_r02_19_180deg_cdf.txt};
    \addlegendimage{color=red, postaction={decoration={ markings, mark=between positions .5 and .5 step .5 with { \fill circle[radius=2pt]; },}, decorate,} }
\addplot[color=teal,postaction={decoration={markings,
      mark=between positions 0 and 1 step 0.1 with { \fill circle[radius=2pt]; },}, decorate,},forget plot] table[x=t,y=exact2]
    {data/r1_5_r2_2_r01_9_r02_19_180deg_cdf.txt};
    \addlegendimage{color=teal, postaction={decoration={ markings, mark=between positions .5 and .5 step .5 with { \fill circle[radius=2pt]; },}, decorate,} }
\addplot[color=orange,postaction={decoration={markings,
      mark=between positions 0 and 1 step 0.1 with { \fill circle[radius=2pt]; },}, decorate,},forget plot] table[x=t,y=sim2]
    {data/r1_5_r2_2_r01_9_r02_19_180deg_cdf.txt};
    \addlegendimage{color=orange, postaction={decoration={ markings, mark=between positions .5 and .5 step .5 with { \fill circle[radius=2pt]; },}, decorate,} }
\addplot[color=blue, dashed] table[x=time,y=sc2_r1]
    {data/all_scs_siso_cdf.txt};
\addplot[color=teal, dashed] table[x=time,y=sc2_r2]
    {data/all_scs_siso_cdf.txt};
\legend{Molecular SIMO response of Rx$_1$ (Comp. Model), Molecular SIMO response of Rx$_1$ (Simulation), Molecular SIMO response of Rx$_2$ (Comp. Model), Molecular SIMO response of Rx$_2$ (Simulation)}  
\end{axis}
\end{tikzpicture} \label{fig:cdfs_scenario2_pi}
} 
\caption{Plots of absorption probability of molecules of Rx$_1$ and Rx$_2$ for Scenario 2.}
\end{figure*}

Similar to the previous scenario, as the separation angle increases, the shadowing on Rx$_2$, which is the farther Rx, decreases. The general trend shows RMS error decrease when separation angle increases, as shown in Fig.~\ref{fig:rms_scenario2}. After separation angle exceeds the no-eclipse angle, separation between receivers becomes relatively distant, therefore co-dependent impact between receivers drops significantly, hence the steady-trending RMS error. The closer Rx is influenced significantly by the release point assumption, therefore the RMS error on Rx$_1$ shows no significant decrease, as shown in Fig.~\ref{fig:rms_scenario2}.


\subsubsection{Scenario 3 -- Same Radii Rxs}
\label{subsubsec:scenario3}
\begin{table}[ht]
\normalsize
\centering
\caption{\normalsize Topology Parameters for Scenario 3}
\begin{tabular}{c||c c c c}
\hline\hline
Parameter & $r_1$ & $r_2$ & $r_{0_1}$ & $r_{0_2}$ \\ \hline
Value & \SI{5}{\micro\meter} & \SI{5}{\micro\meter} & \SI{9}{\micro\meter} & \SI{22}{\micro\meter}\\ \hline\hline
\end{tabular}
\label{table:sc3}
\end{table}

This configuration offers a topology of Rxs of the same radii. Since the {tangent cone} of the closer Rx span a larger angle centered by Tx, the closer Rx still shadows the farther Rx when the separation angle $\phi$ is close or equal to $0$. This again prevents the farther Rx to receive molecules significantly when it is eclipsed by the closer Rx.

\begin{figure}[ht]
\centering
\begin{tikzpicture}
\begin{axis}[ width=\linewidth, height=.5\linewidth,
    xlabel={Separation angle $(\deg)$}, ylabel={RMS error},
    ymin=0, grid, every axis plot/.append style={very thick},
    xtick={0, 60, 90, 120, 150, 180},
    extra x ticks={33.74, 46.88},
    extra x tick labels={$\varphi_1$,$\varphi_2$},
    legend style={at={(0.99,0.99)},anchor=north east,
    nodes={scale=0.38, transform shape}}]
\addplot[color=blue,mark=o] table[x=angles,y=rms_exact1]
    {data/r1_5_r2_5_perf.txt};
\addplot[color=red,mark=o] table[x=angles,y=rms_exact2]
    {data/r1_5_r2_5_perf.txt};
\addplot[color=blue,mark=square*,fill=blue,only marks]
    table[x=angles,y=rms_exact1] {data/r1_5_r2_5_halfEclipse.txt};
\addplot[color=red,mark=square*,fill=red,only marks]
    table[x=angles,y=rms_exact2] {data/r1_5_r2_5_halfEclipse.txt};
\addplot[color=blue,mark=triangle*,fill=blue,only marks]
    table[x=angles,y=rms_exact1] {data/r1_5_r2_5_noEclipse.txt};
\addplot[color=red,mark=triangle*,fill=red,only marks]
    table[x=angles,y=rms_exact2] {data/r1_5_r2_5_noEclipse.txt};
\draw (33.749,0) -- (33.749,3e-2);
\draw (46.886,0) -- (46.886,3e-2);
\legend{RMS error of Rx$_1$ (Comp. Model), RMS error of Rx$_2$ (Comp. Model), RMS error of Rx$_1$ at half eclipse angle, RMS error of Rx$_2$ at half eclipse angle,RMS error of Rx$_1$ at no eclipse angle,RMS error of Rx$_1$ at no eclipse angle}    
\end{axis}
\end{tikzpicture}%
\caption{Plots of RMS error of Rx$_1$ and Rx$_2$ for Scenario 3, with variable angular separation, $\varphi_1 = 33.75\ \SI{}{\deg}$, $\varphi_2 = 46.88\ \SI{}{\deg}$.}
\label{fig:rms_scenario3}
\end{figure}

\begin{figure}[ht]
\centering
\subcaptionbox{Visualization of the topology for Scenario 3.}%
[.45\linewidth]{
\includegraphics[width=\linewidth]{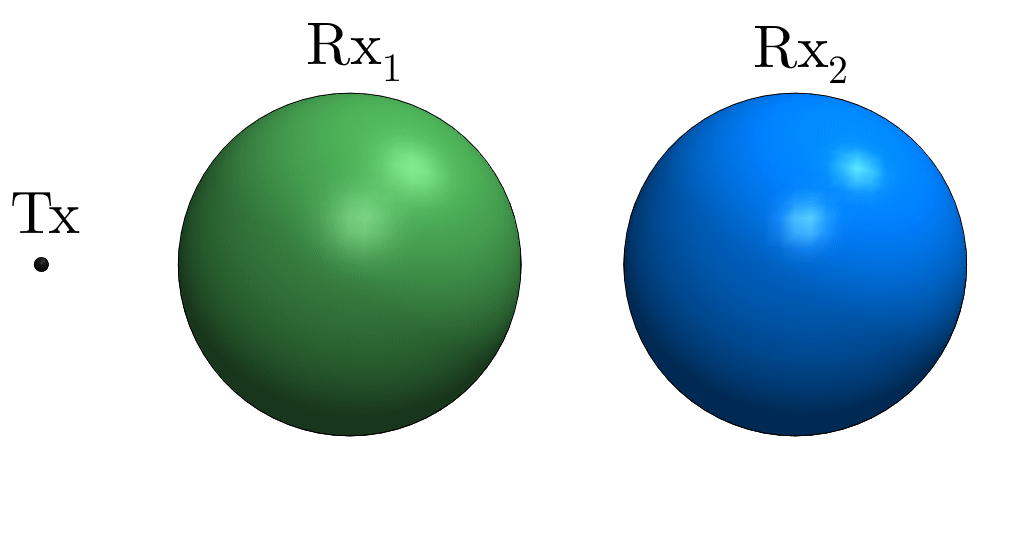}
\label{fig:scenario3_topology}
} 
\hfill
\subcaptionbox{Visualization of the topology for Multi-Receiver Scenario with 3 Rxs.}%
[.45\linewidth]{
\includegraphics[width=\linewidth]{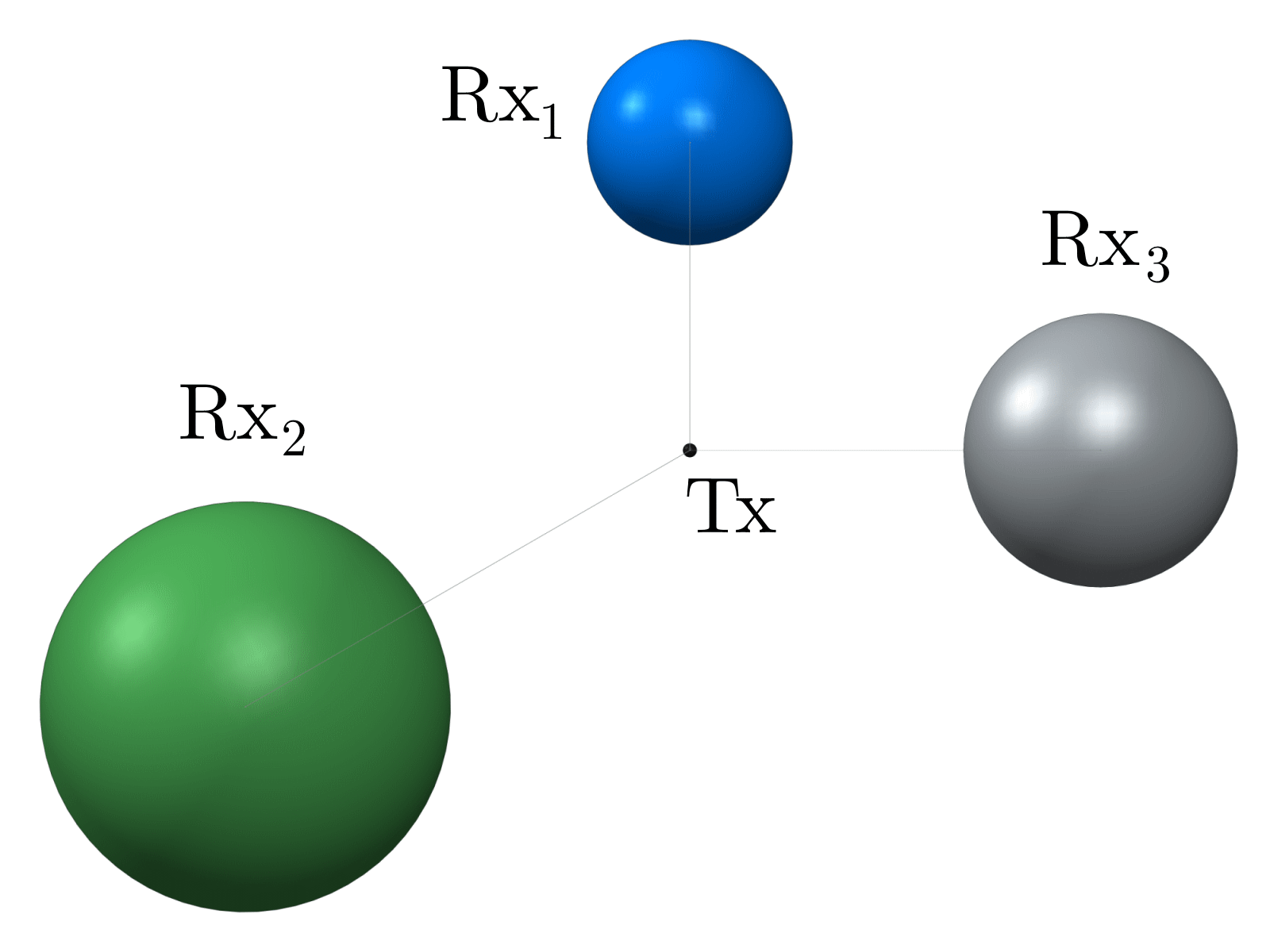}
\label{fig:multi_rx_topology}
} 
\caption{Visualizations of the topologies for Scenario 3 and Multi-Receiver Scenario.}
\end{figure}

\begin{figure*}[ht]
\centering
\subcaptionbox{Plots of absorption probability of molecules of Rx$_1$ and Rx$_2$ for Scenario 3 placed with $\varphi_1 = \arcsin {\frac{r_1}{r_{0_1}}}\ \SI{}{\radian}$ angular separation.}%
[.32\textwidth]{
\begin{tikzpicture}
\begin{axis}[width=\linewidth, height=1.2\linewidth,
    xlabel={Time $(s)$}, ylabel={Absorption probability of molecules}, ymin=0, ymax=0.55, xmin=0, xmax=5, grid,
    every axis plot/.append style={very thick},
    y tick label style={ /pgf/number format/.cd,
    fixed, fixed zerofill, precision=2},
    legend style={at={(0.98,0.4)},anchor=east,
    nodes={scale=0.4, transform shape}}]
\addplot[color=blue,postaction={decoration={markings,
      mark=between positions 0 and 1 step 0.1 with { \fill circle[radius=2pt]; },}, decorate,},forget plot] table[x=t,y=exact1]
    {data/r1_5_r2_5_r01_9_r02_22_halfEclipse_cdf.txt};
    \addlegendimage{color=blue, postaction={decoration={ markings, mark=between positions .5 and .5 step .5 with { \fill circle[radius=2pt]; },}, decorate,} }
\addplot[color=red,postaction={decoration={markings,
      mark=between positions 0 and 1 step 0.1 with { \fill circle[radius=2pt]; },}, decorate,},forget plot] table[x=t,y=sim1]
    {data/r1_5_r2_5_r01_9_r02_22_halfEclipse_cdf.txt};
    \addlegendimage{color=red, postaction={decoration={ markings, mark=between positions .5 and .5 step .5 with { \fill circle[radius=2pt]; },}, decorate,} }
\addplot[color=teal,postaction={decoration={markings,
      mark=between positions 0 and 1 step 0.1 with { \fill circle[radius=2pt]; },}, decorate,},forget plot] table[x=t,y=exact2]
    {data/r1_5_r2_5_r01_9_r02_22_halfEclipse_cdf.txt};
    \addlegendimage{color=teal, postaction={decoration={ markings, mark=between positions .5 and .5 step .5 with { \fill circle[radius=2pt]; },}, decorate,} }
\addplot[color=orange,postaction={decoration={markings,
      mark=between positions 0 and 1 step 0.1 with { \fill circle[radius=2pt]; },}, decorate,},forget plot] table[x=t,y=sim2]
    {data/r1_5_r2_5_r01_9_r02_22_halfEclipse_cdf.txt};
    \addlegendimage{color=orange, postaction={decoration={ markings, mark=between positions .5 and .5 step .5 with { \fill circle[radius=2pt]; },}, decorate,} }
\addplot[color=blue, dashed] table[x=time,y=sc3_r1]
    {data/all_scs_siso_cdf.txt};
\addplot[color=teal, dashed] table[x=time,y=sc3_r2]
    {data/all_scs_siso_cdf.txt};
\legend{Molecular SIMO response of Rx$_1$ (Comp. Model), Molecular SIMO response of Rx$_1$ (Simulation), Molecular SIMO response of Rx$_2$ (Comp. Model), Molecular SIMO response of Rx$_2$ (Simulation)}
\end{axis}
\end{tikzpicture} \label{fig_cdfs_scenario3_half}
} 
\subcaptionbox{Plots of absorption probability of molecules of Rx$_1$ and Rx$_2$ for Scenario 3 placed with $\frac{\pi}{2}\ \SI{}{\radian}$ angular separation.}%
[.32\textwidth]{
\begin{tikzpicture}
\begin{axis}[width=\linewidth, height=1.2\linewidth,
    xlabel={Time $(s)$}, ylabel={Absorption probability of molecules}, ymin=0, ymax=0.55, xmin=0, xmax=5, grid,
    every axis plot/.append style={very thick},
    y tick label style={ /pgf/number format/.cd,
    fixed, fixed zerofill, precision=2},
    legend style={at={(0.98,0.4)},anchor=east,
    nodes={scale=0.4, transform shape}}]
\addplot[color=blue,postaction={decoration={markings,
      mark=between positions 0 and 1 step 0.1 with { \fill circle[radius=2pt]; },}, decorate,},forget plot] table[x=t,y=exact1]
    {data/r1_5_r2_5_r01_9_r02_22_90deg_cdf.txt};
    \addlegendimage{color=blue, postaction={decoration={ markings, mark=between positions .5 and .5 step .5 with { \fill circle[radius=2pt]; },}, decorate,} }
\addplot[color=red,postaction={decoration={markings,
      mark=between positions 0 and 1 step 0.1 with { \fill circle[radius=2pt]; },}, decorate,},forget plot] table[x=t,y=sim1]
    {data/r1_5_r2_5_r01_9_r02_22_90deg_cdf.txt};
    \addlegendimage{color=red, postaction={decoration={ markings, mark=between positions .5 and .5 step .5 with { \fill circle[radius=2pt]; },}, decorate,} }
\addplot[color=teal,postaction={decoration={markings,
      mark=between positions 0 and 1 step 0.1 with { \fill circle[radius=2pt]; },}, decorate,},forget plot] table[x=t,y=exact2]
    {data/r1_5_r2_5_r01_9_r02_22_90deg_cdf.txt};
    \addlegendimage{color=teal, postaction={decoration={ markings, mark=between positions .5 and .5 step .5 with { \fill circle[radius=2pt]; },}, decorate,} }
\addplot[color=orange,postaction={decoration={markings,
      mark=between positions 0 and 1 step 0.1 with { \fill circle[radius=2pt]; },}, decorate,},forget plot] table[x=t,y=sim2]
    {data/r1_5_r2_5_r01_9_r02_22_90deg_cdf.txt};
    \addlegendimage{color=orange, postaction={decoration={ markings, mark=between positions .5 and .5 step .5 with { \fill circle[radius=2pt]; },}, decorate,} }
\addplot[color=blue, dashed] table[x=time,y=sc3_r1]
    {data/all_scs_siso_cdf.txt};
\addplot[color=teal, dashed] table[x=time,y=sc3_r2]
    {data/all_scs_siso_cdf.txt};
\legend{Molecular SIMO response of Rx$_1$ (Comp. Model), Molecular SIMO response of Rx$_1$ (Simulation), Molecular SIMO response of Rx$_2$ (Comp. Model), Molecular SIMO response of Rx$_2$ (Simulation)}
\end{axis}
\end{tikzpicture} \label{fig:cdfs_scenario3_pi_2}
} 
\subcaptionbox{Plots of absorption probability of molecules of Rx$_1$ and Rx$_2$ for Scenario 3 placed with $\pi\ \SI{}{\radian}$ angular separation.}%
[.32\textwidth]{
\begin{tikzpicture}
\begin{axis}[width=\linewidth, height=1.2\linewidth,
    xlabel={Time $(s)$}, ylabel={Absorption probability of molecules}, ymin=0, ymax=0.55, xmin=0, xmax=5, grid,
    every axis plot/.append style={very thick},
    y tick label style={ /pgf/number format/.cd,
    fixed, fixed zerofill, precision=2},
    legend style={at={(0.98,0.4)},anchor=east,
    nodes={scale=0.4, transform shape}}]
\addplot[color=blue,postaction={decoration={markings,
      mark=between positions 0 and 1 step 0.1 with { \fill circle[radius=2pt]; },}, decorate,},forget plot] table[x=t,y=exact1] {data/r1_5_r2_5_r01_9_r02_22_180deg_cdf.txt};
      \addlegendimage{color=blue, postaction={decoration={ markings, mark=between positions .5 and .5 step .5 with { \fill circle[radius=2pt]; },}, decorate,} }
\addplot[color=red,postaction={decoration={markings,
      mark=between positions 0 and 1 step 0.1 with { \fill circle[radius=2pt]; },}, decorate,},forget plot] table[x=t,y=sim1]
    {data/r1_5_r2_5_r01_9_r02_22_180deg_cdf.txt};
    \addlegendimage{color=red, postaction={decoration={ markings, mark=between positions .5 and .5 step .5 with { \fill circle[radius=2pt]; },}, decorate,} }
\addplot[color=teal,postaction={decoration={markings,
      mark=between positions 0 and 1 step 0.1 with { \fill circle[radius=2pt]; },}, decorate,},forget plot] table[x=t,y=exact2]
    {data/r1_5_r2_5_r01_9_r02_22_180deg_cdf.txt};
    \addlegendimage{color=teal, postaction={decoration={ markings, mark=between positions .5 and .5 step .5 with { \fill circle[radius=2pt]; },}, decorate,} }
\addplot[color=orange,postaction={decoration={markings,
      mark=between positions 0 and 1 step 0.1 with { \fill circle[radius=2pt]; },}, decorate,},forget plot] table[x=t,y=sim2]
    {data/r1_5_r2_5_r01_9_r02_22_180deg_cdf.txt};
    \addlegendimage{color=orange, postaction={decoration={ markings, mark=between positions .5 and .5 step .5 with { \fill circle[radius=2pt]; },}, decorate,} }
\addplot[color=blue, dashed] table[x=time,y=sc3_r1]
    {data/all_scs_siso_cdf.txt};
\addplot[color=teal, dashed] table[x=time,y=sc3_r2]
    {data/all_scs_siso_cdf.txt};
\legend{Molecular SIMO response of Rx$_1$ (Comp. Model), Molecular SIMO response of Rx$_1$ (Simulation), Molecular SIMO response of Rx$_2$ (Comp. Model), Molecular SIMO response of Rx$_2$ (Simulation)}
\end{axis}
\end{tikzpicture} \label{fig:cdfs_scenario3_pi}
} 
\caption{Plots of absorption probability of molecules of Rx$_1$ and Rx$_2$ for Scenario 3.}
\end{figure*}

In this case, farther Rx escapes the shadowing earlier than the previous scenario as the separation angle increases due to larger radius. Therefore, the performance increase of Rx$_2$ has a steeper trend in terms of RMS error. Co-dependence between receivers drops eventually after the no-eclipse angle, hence the steady-trending RMS error on both receivers, as shown in Fig.~\ref{fig:rms_scenario3}. The farther Rx is significantly far away from Tx, therefore it absorbs fairly small number of molecules compared to the closer Rx. Consequently, the placement of the release point towards Rx$_2$ significantly impacts the absorption probability of molecules, and consequently the RMS error.

As expected, the shadowing phenomenon impacts the performance of the channel response modeling significantly. The recursive model provides well-fit channel response estimation with significantly small error. RMS error decreases monotonically as the shadowing diminishes. One of the reasons for error increase is based on the release point assumption. As mentioned before, the release point from an Rx is assumed {to be a point within Rx towards Tx, to be calculated via \eqref{eq:siso_center_of_mass}}. For close Rx pairs, as the separation angle between Rxs goes to $0$, the release point to Rxs with relatively small radii is drastically influenced. On the other hand, the release point to Rxs with relatively large radii is not impacted in a comparable significance. This causes Rxs with relatively large radii to be robust against considerable margin of error in the release point assumption. Considering the RMS error order of magnitudes, the errors that are encountered are significantly small. Provided that the distance between the release points and the centers of regarding Rxs is sufficiently large, the recursive model fits to the simulation results for the channel response of both Rxs with sufficiently low RMS error.

\subsection{Multi-Receiver Scenario with More Than Two Receivers}
\label{subsection:Multi_rx_scenario}
\begin{table}[ht]
\normalsize
\centering
\caption{\normalsize Topology Parameters Multi-Receiver Scenario}
\begin{tabular}{c||c c c c c}
\hline\hline
Parameter & $r_1$ & $r_2$ & $r_3$ & $r_{0_1}$ & $r_{0_2}$\\ \hline
Value & \SI{3}{\micro\meter} & \SI{6}{\micro\meter} & \SI{4}{\micro\meter} & \SI{9}{\micro\meter} & \SI{15}{\micro\meter}\\ \hline
Parameter & $r_{0_3}$ & $\phi_1$ & $\phi_2$ & $\phi_3$ \\ \hline
Value & \SI{12}{\micro\meter} & $0\ \SI{}{\radian}$ & $\frac{2\pi}{3}\ \SI{}{\radian}$ & $\frac{3\pi}{2}\ \SI{}{\radian}$ \\ \hline\hline
\end{tabular}
\label{table:multi_rx}
\end{table}

To emphasize the applicability of the proposed model to the molecular systems with multiple Rxs, more than two Rxs in particular, a molecular SIMO system with three Rxs is investigated with respect to the absorption probabilities of molecules. The molecular SIMO system consists of three Rxs, with radii of $\SI{3}{\micro\meter}$, $\SI{6}{\micro\meter}$, and $\SI{4}{\micro\meter}$, respectively. The center-to-center distances $r_{0_1}$, $r_{0_2}$, and $r_{0_3}$ are $\SI{9}{\micro\meter}$, $\SI{15}{\micro\meter}$, and $\SI{12}{\micro\meter}$, respectively. The separation angle between Rx$_1$ and Rx$_2$ is $\phi_{1-2} = \phi_2 - \phi_1 = \frac{2\pi}{3} \SI{}{\radian}$. Similarly, the separation angle between Rx$_1$ and Rx$_3$ is $\phi_{1-3} = \phi_3 - \phi_1 = \frac{\pi}{2} \SI{}{\radian}$. And finally, the separation angle between Rx$_2$ and Rx$_3$ is $\phi_{2-3} = \phi_3 - \phi_2 = \frac{5\pi}{6} \SI{}{\radian}$.

\begin{figure}[ht]
\centering
\begin{tikzpicture}
\begin{axis}[width=\linewidth, height=.7\linewidth,
    xlabel={Time $(s)$}, ylabel={Absorption probability of molecules}, ymin=0, ymax=0.3, xmin=0, xmax=3, grid,
    every axis plot/.append style={very thick},
    y tick label style={/pgf/number format/.cd,
    fixed, fixed zerofill, precision=2 },
    legend style = {at={(0.98,0.02)}, anchor=south east,
    nodes={scale=0.5, transform shape}}]
\addplot[color=blue, postaction={ decoration={ markings,
      mark=between positions 0 and 1 step 0.1 with { \fill circle[radius=2pt]; },}, decorate,},forget plot] table[x=t,y=exact1] {data/r1_3_r2_6_r3_4_r01_9_r02_15_r03_12_ang1_0deg_ang2_120deg_ang3_270deg_cdf.txt};
      \addlegendimage{color=blue, postaction={decoration={ markings, mark=between positions .5 and .5 step .5 with { \fill circle[radius=2pt]; },}, decorate,} }
\addplot[color=red,postaction={ decoration={ markings,
      mark=between positions 0 and 1 step 0.1 with { \fill circle[radius=2pt]; },}, decorate,},forget plot] table[x=t,y=sim1] {data/r1_3_r2_6_r3_4_r01_9_r02_15_r03_12_ang1_0deg_ang2_120deg_ang3_270deg_cdf.txt};
      \addlegendimage{color=red, postaction={decoration={ markings, mark=between positions .5 and .5 step .5 with { \fill circle[radius=2pt]; },}, decorate,} }
\addplot[color=teal,postaction={ decoration={ markings,
      mark=between positions 0 and 1 step 0.1 with { \fill circle[radius=2pt]; },}, decorate,},forget plot] table[x=t,y=exact2] {data/r1_3_r2_6_r3_4_r01_9_r02_15_r03_12_ang1_0deg_ang2_120deg_ang3_270deg_cdf.txt};
      \addlegendimage{color=teal, postaction={decoration={ markings, mark=between positions .5 and .5 step .5 with { \fill circle[radius=2pt]; },}, decorate,} }
\addplot[color=orange,postaction={ decoration={ markings,
      mark=between positions 0 and 1 step 0.1 with { \fill circle[radius=2pt]; },}, decorate,},forget plot] table[x=t,y=sim2]{data/r1_3_r2_6_r3_4_r01_9_r02_15_r03_12_ang1_0deg_ang2_120deg_ang3_270deg_cdf.txt};
      \addlegendimage{color=orange, postaction={decoration={ markings, mark=between positions .5 and .5 step .5 with { \fill circle[radius=2pt]; },}, decorate,} }
\addplot[color=brown,postaction={ decoration={ markings,
      mark=between positions 0 and 1 step 0.1 with { \fill circle[radius=2pt]; },}, decorate,},forget plot] table[x=t,y=exact3]{data/r1_3_r2_6_r3_4_r01_9_r02_15_r03_12_ang1_0deg_ang2_120deg_ang3_270deg_cdf.txt};
      \addlegendimage{color=brown, postaction={decoration={ markings, mark=between positions .5 and .5 step .5 with { \fill circle[radius=2pt]; },}, decorate,} }
\addplot[color=purple,postaction={ decoration={ markings,
      mark=between positions 0 and 1 step 0.1 with { \fill circle[radius=2pt]; },}, decorate,},forget plot] table[x=t,y=sim3] {data/r1_3_r2_6_r3_4_r01_9_r02_15_r03_12_ang1_0deg_ang2_120deg_ang3_270deg_cdf.txt};
      \addlegendimage{color=purple, postaction={decoration={ markings, mark=between positions .5 and .5 step .5 with { \fill circle[radius=2pt]; },}, decorate,} }
\addplot[color=blue, dashed] table[x=time,y=rx1]
    {data/large_sc_siso_cdf.txt};
\addplot[color=teal, dashed] table[x=time,y=rx2]
    {data/large_sc_siso_cdf.txt};
\addplot[color=brown, dashed] table[x=time,y=rx3]
    {data/large_sc_siso_cdf.txt};
\legend{Molecular SIMO response of Rx$_1$ (Comp. Model), Molecular SIMO response of Rx$_1$ (Simulation), Molecular SIMO response of Rx$_2$ (Comp. Model), Molecular SIMO response of Rx$_2$ (Simulation), Molecular SIMO response of Rx$_3$ (Comp. Model), Molecular SIMO response of Rx$_3$ (Simulation), Molecular SISO response of Rx$_1$ (Theoretical), Molecular SISO response of Rx$_2$ (Theoretical), Molecular SISO response of Rx$_3$ (Theoretical)} 
\end{axis}
\end{tikzpicture}%
\caption{Plots of absorption probability of molecules of Rx$_1$, Rx$_2$, and Rx$_3$ for multi-receiver scenario.}
\label{fig:multi_rx_cdfs}
\end{figure}

The dependence among the Rxs is considered when the recursive model is applied to the molecular SIMO system to acquire the absorption probabilities of molecules by Rxs. The results show that for crowded molecular networks, such as the proposed molecular multi-receiver system, the recursive model offers a well-fit channel behavior estimation for each receiver. Note that Rx sets whose Rxs are placed closer to each other such that they interfere with one another significantly may deviate from the simulation-based results more than expected. However, it is safe to say that the extension of the recursive model, and naturally the simplified model, towards the crowded molecular SIMO systems is justified with the channel response estimation proposed in Fig.~\ref{fig:multi_rx_cdfs}.

\subsection{Model Correlation Analysis}
\label{subsection:ModelCorrAnalysis}
The recursive model has high complexity, and requires higher computational power. {The derivation steps can be modeled in a discrete manner in a computer environment, therefore, time is a variable to be discretely modeled. Therefore, in such computations,} the shorter the time step $\Delta t$, the higher the computational requirement is. Although it provides a higher-accuracy model for molecular SIMO channel response, it is analytically derived as an infinite sum. Coefficients of further terms tend to go to $0$, however, it still requires significant amount of calculations.

\begin{figure*}[ht]
\centering
\subcaptionbox{Plots of absorption probability of molecules of Rx$_1$ and Rx$_2$ for Scenario 1 placed with $\frac{\pi}{2}\ \SI{}{\radian}$ angular separation. \label{fig:cdfs_sc1_correlation_pi_2}} 
[.24\textwidth]{
\begin{tikzpicture}
\begin{axis}[width=\linewidth, height=1.4\linewidth,
    xlabel={Time $(s)$}, ylabel={Absorption probability of molecules}, ymin=0, ymax=0.4, xmin=0, xmax=5, grid,
    every axis plot/.append style={very thick},
    y tick label style={/pgf/number format/.cd,
    fixed, fixed zerofill, precision=2},
    legend style = {at={(0.99,0.995)}, anchor=north east,
    nodes={scale=0.25, transform shape}}]
\addplot[color=blue, postaction={decoration={ markings,
      mark=between positions 0 and 1 step 0.1 with { \fill circle[radius=2pt]; },}, decorate,}, forget plot] table[x=t,y=exact1]
    {data/r1_2_r2_5_r01_6_r02_16_90deg_cdf.txt};
    \addlegendimage{color=blue, postaction={decoration={ markings,
      mark=between positions .5 and .5 step .5 with { \fill circle[radius=2pt]; },}, decorate,} }
\addplot[color=red,postaction={decoration={ markings,
      mark=between positions 0 and 1 step 0.1 with { \fill circle[radius=2pt]; },}, decorate,}, forget plot] table[x=t,y=appr1]
    {data/r1_2_r2_5_r01_6_r02_16_90deg_cdf.txt};
    \addlegendimage{color=red, postaction={decoration={ markings,
      mark=between positions .5 and .5 step .5 with { \fill circle[radius=2pt]; },}, decorate,} }
\addplot[color=teal,postaction={decoration={ markings,
      mark=between positions 0 and 1 step 0.1 with { \fill circle[radius=2pt]; },}, decorate,}, forget plot] table[x=t,y=exact2]
    {data/r1_2_r2_5_r01_6_r02_16_90deg_cdf.txt};
    \addlegendimage{color=teal, postaction={decoration={ markings,
      mark=between positions .5 and .5 step .5 with { \fill circle[radius=2pt]; },}, decorate,} }
\addplot[color=orange,postaction={decoration={ markings,
      mark=between positions 0 and 1 step 0.1 with { \fill circle[radius=2pt]; },}, decorate,}, forget plot] table[x=t,y=appr2]
    {data/r1_2_r2_5_r01_6_r02_16_90deg_cdf.txt};
    \addlegendimage{color=orange, postaction={decoration={ markings, mark=between positions .5 and .5 step .5 with { \fill circle[radius=2pt]; },}, decorate,} }
\legend{Molecular SIMO response of Rx$_1$ (Comp. Model), Molecular SIMO response of Rx$_1$ (Appr. Model), Molecular SIMO response of Rx$_2$ (Comp. Model), Molecular SIMO response of Rx$_2$ (Appr. Model)}
\end{axis}
\end{tikzpicture}
} 
\subcaptionbox{Plots of absorption probability of molecules of Rx$_1$ and Rx$_2$ for Scenario 1 placed with $\varphi_1 = \arcsin {\frac{r_1}{r_{0_1}}}\ \SI{}{\radian}$  angular separation. \label{fig:cdfs_sc1_correlation_half}}%
[.24\textwidth]{
\begin{tikzpicture}
\begin{axis}[width=\linewidth, height=1.4\linewidth,
xlabel={Time $(s)$}, ylabel={Absorption probability of molecules}, ymin=0, ymax=0.4, xmin=0, xmax=5, grid,
every axis plot/.append style={very thick},
y tick label style={/pgf/number format/.cd,
fixed, fixed zerofill, precision=2}, 
legend style={at={(0.99,0.995)},anchor=north east,
nodes={scale=0.25, transform shape}}]
\addplot[color=blue,postaction={decoration={ markings,
mark=between positions 0 and 1 step 0.1 with { \fill circle[radius=2pt]; },}, decorate,}, forget plot] table[x=t,y=exact1]
{data/r1_2_r2_5_r01_6_r02_16_halfEclipse_cdf.txt};
\addlegendimage{color=blue, postaction={decoration={ markings,
mark=between positions .5 and .5 step .5 with { \fill circle[radius=2pt]; },}, decorate,} }
\addplot[color=red,postaction={decoration={ markings,
mark=between positions 0 and 1 step 0.1 with { \fill circle[radius=2pt]; },}, decorate,}, forget plot] table[x=t,y=appr1]
{data/r1_2_r2_5_r01_6_r02_16_halfEclipse_cdf.txt};
\addlegendimage{color=red, postaction={decoration={ markings,
mark=between positions .5 and .5 step .5 with { \fill circle[radius=2pt]; },}, decorate,} }
\addplot[color=teal,postaction={decoration={ markings,
mark=between positions 0 and 1 step 0.1 with { \fill circle[radius=2pt]; },}, decorate,}, forget plot] table[x=t,y=exact2]
{data/r1_2_r2_5_r01_6_r02_16_halfEclipse_cdf.txt};
\addlegendimage{color=teal, postaction={decoration={ markings,
mark=between positions .5 and .5 step .5 with { \fill circle[radius=2pt]; },}, decorate,} }
\addplot[color=orange,postaction={decoration={ markings,
mark=between positions 0 and 1 step 0.1 with { \fill circle[radius=2pt]; },}, decorate,},forget plot] table[x=t,y=appr2]
{data/r1_2_r2_5_r01_6_r02_16_halfEclipse_cdf.txt};
\addlegendimage{color=orange, postaction={decoration={ markings,
mark=between positions .5 and .5 step .5 with { \fill circle[radius=2pt]; },}, decorate,} }
\legend{Molecular SIMO response of Rx$_1$ (Comp. Model), Molecular SIMO response of Rx$_1$ (Appr. Model), Molecular SIMO response of Rx$_2$ (Comp. Model), Molecular SIMO response of Rx$_2$ (Appr. Model)} 
\end{axis}
\end{tikzpicture} 
} 
\subcaptionbox{Plots of absorption probability of molecules of Rx$_1$ and Rx$_2$ for Scenario 3 placed with $\frac{\pi}{2}\ \SI{}{\radian}$ angular separation. \label{fig:cdfs_sc3_correlation_pi_2}}%
[.24\textwidth]{
\begin{tikzpicture}
\begin{axis}[width=\linewidth, height=1.4\linewidth,
    xlabel={Time $(s)$}, ylabel={Absorption probability of molecules}, ymin=0, ymax=0.65, xmin=0, xmax=5, grid,
    every axis plot/.append style={very thick},
    y tick label style={/pgf/number format/.cd,
    fixed, fixed zerofill, precision=2},
    legend style={at={(0.99,0.995)},anchor=north east,
    nodes={scale=0.25, transform shape}}]
\addplot[color=blue,postaction={decoration={ markings,
      mark=between positions 0 and 1 step 0.1 with { \fill circle[radius=2pt]; },}, decorate,},forget plot] table[x=t,y=exact1]
    {data/r1_5_r2_5_r01_9_r02_22_90deg_cdf.txt};
    \addlegendimage{color=blue, postaction={decoration={ markings,
      mark=between positions .5 and .5 step .5 with { \fill circle[radius=2pt]; },}, decorate,} }
\addplot[color=red,postaction={decoration={ markings,
      mark=between positions 0 and 1 step 0.1 with { \fill circle[radius=2pt]; },}, decorate,},forget plot]  table[x=t,y=appr1]
    {data/r1_5_r2_5_r01_9_r02_22_90deg_cdf.txt};
    \addlegendimage{color=red, postaction={decoration={ markings,
      mark=between positions .5 and .5 step .5 with { \fill circle[radius=2pt]; },}, decorate,} }
\addplot[color=teal,postaction={decoration={ markings,
      mark=between positions 0 and 1 step 0.1 with { \fill circle[radius=2pt]; },}, decorate,},forget plot]  table[x=t,y=exact2]
    {data/r1_5_r2_5_r01_9_r02_22_90deg_cdf.txt};
    \addlegendimage{color=teal, postaction={decoration={ markings,
      mark=between positions .5 and .5 step .5 with { \fill circle[radius=2pt]; },}, decorate,} }
\addplot[color=orange,postaction={decoration={ markings,
      mark=between positions 0 and 1 step 0.1 with { \fill circle[radius=2pt]; },}, decorate,},forget plot]  table[x=t,y=appr2]
    {data/r1_5_r2_5_r01_9_r02_22_90deg_cdf.txt};
    \addlegendimage{color=orange, postaction={decoration={ markings, mark=between positions .5 and .5 step .5 with { \fill circle[radius=2pt]; },}, decorate,} }
\legend{Molecular SIMO response of Rx$_1$ (Comp. Model), Molecular SIMO response of Rx$_1$ (Appr. Model), Molecular SIMO response of Rx$_2$ (Comp. Model), Molecular SIMO response of Rx$_2$ (Appr. Model)} 
\end{axis}
\end{tikzpicture}
} 
\subcaptionbox{Plots of absorption probability of molecules of Rx$_1$ and Rx$_2$ for Scenario 3 placed with $\varphi_1 = \arcsin {\frac{r_1}{r_{0_1}}}\ \SI{}{\radian}$ angular separation. \label{fig:cdfs_sc3_correlation_half}}%
[.24\textwidth] {
\begin{tikzpicture}
\begin{axis}[width=\linewidth, height=1.4\linewidth,
    xlabel={Time $(s)$}, ylabel={Absorption probability of molecules}, ymin=0, ymax=0.65, xmin=0, xmax=5, grid,
    every axis plot/.append style={very thick},
    y tick label style={/pgf/number format/.cd,
    fixed, fixed zerofill, precision=2}, 
    legend style={at={(0.99,0.995)},anchor=north east,
    nodes={scale=0.25, transform shape}}]
\addplot[color=blue,postaction={decoration={ markings,
      mark=between positions 0 and 1 step 0.1 with { \fill circle[radius=2pt]; },}, decorate,},forget plot] table[x=t,y=exact1]
    {data/r1_5_r2_5_r01_9_r02_22_halfEclipse_cdf.txt};
    \addlegendimage{color=blue, postaction={decoration={ markings, mark=between positions .5 and .5 step .5 with { \fill circle[radius=2pt]; },}, decorate,} }
\addplot[color=red,postaction={decoration={ markings,
      mark=between positions 0 and 1 step 0.1 with { \fill circle[radius=2pt]; },}, decorate,},forget plot] table[x=t,y=appr1]
    {data/r1_5_r2_5_r01_9_r02_22_halfEclipse_cdf.txt};
    \addlegendimage{color=red, postaction={decoration={ markings, mark=between positions .5 and .5 step .5 with { \fill circle[radius=2pt]; },}, decorate,} }
\addplot[color=teal,postaction={decoration={ markings,
      mark=between positions 0 and 1 step 0.1 with { \fill circle[radius=2pt]; },}, decorate,},forget plot] table[x=t,y=exact2]
    {data/r1_5_r2_5_r01_9_r02_22_halfEclipse_cdf.txt};
    \addlegendimage{color=teal, postaction={decoration={ markings, mark=between positions .5 and .5 step .5 with { \fill circle[radius=2pt]; },}, decorate,} }
\addplot[color=orange,postaction={decoration={ markings,
      mark=between positions 0 and 1 step 0.1 with { \fill circle[radius=2pt]; },}, decorate,},forget plot] table[x=t,y=appr2]
    {data/r1_5_r2_5_r01_9_r02_22_halfEclipse_cdf.txt};
    \addlegendimage{color=orange, postaction={decoration={ markings, mark=between positions .5 and .5 step .5 with { \fill circle[radius=2pt]; },}, decorate,} }
\legend{Molecular SIMO response of Rx$_1$ (Comp. Model), Molecular SIMO response of Rx$_1$ (Appr. Model), Molecular SIMO response of Rx$_2$ (Comp. Model), Molecular SIMO response of Rx$_2$ (Appr. Model)} 
\end{axis}
\end{tikzpicture}
} 
\caption{Plots of absorption probability of molecules of Rx$_1$ and Rx$_2$ for Scenario 1 and Scenario 3.}
\end{figure*}

Based on the assumption on the further terms going to $0$, the simplified model ignores these terms to provide a relatively-low-complexity molecular SIMO modeling. The terms which correspond to the effect of the molecules \textit{theoretically} absorbed by multiple times, become significant in certain cases such as the topologies of receivers close to each other. 

The comparison between the recursive model and the simplified model exhibits significant correlation between the models, as shown in Fig.~\ref{fig:cdfs_sc1_correlation_half}, \ref{fig:cdfs_sc1_correlation_pi_2}, \ref{fig:cdfs_sc3_correlation_half}, and \ref{fig:cdfs_sc3_correlation_pi_2}. In given topologies, smaller separation angles correspond to closer Rx placements, which require higher terms to be taken into account. In cases where the separation angle is smaller than no-eclipse angle, the simplified model slightly deviates from the recursive model. In other cases where the angular separation is significant and the impact among Rxs is sufficiently weakened, the models correlate quite eloquently. {Our computer-based simulations on the significance of higher terms showed that for the molecular SIMO systems in which the receivers are placed approximately two to three times of their radii or more from each other, the mentioned receivers affect one another less and less. Although this effect depends on multitude of spatial parameters, we can roughly state that the deviation between the output of the recursive model and that of the simplified model is bound to be less than $5\%$ as time goes to infinity, when the condition above is satisfied. As the distance between the receiver increases, the effect radically continues to decrease, and the outputs of both models converge.}

In Subsection~\ref{subsection:angular_analysis}, the performance of the recursive model has been evaluated. Considering both analyses, the performance of the simplified model compared to the computer simulation results is significant, and the model results can be said to fit to the molecular SIMO channel response generated by the computer simulations.



\section{Conclusion}
\label{section:Conclusion}
In this study, channel modeling for molecular SIMO topologies is presented. Analytical derivations for molecular SISO channel are well-conducted. However, this study points to the absence of the proper analytical expressions derived for multiple-receiver topologies. The recursive model takes the potential future impacts of the previously-absorbed molecules into account, recursively. The mentioned future impacts are removed from the upper-bound channel responses. The recursive expressions are used to derive the comprehensive analytical channel response expressions. Comparing to the computer simulation-based {absorption probability} of molecules over time, the recursive model shows promising performance in terms of correlation and RMS error metrics. The molecular SIMO topologies can be parameterized by distances from the Tx, radii of receivers, and the separation angle among receivers with respect to Tx. As expected, in such cases where the separation angle is small, Rxs are close to each other and affect each other significantly. In these cases, the error of the model compared to the computer simulation data increases slightly in terms of RMS error. This is mainly because of the rough release point assumption. As the separation angle increases, the displacement among receivers increases, therefore the influence among receivers reduces. This enables the models to predict the channel response more accurately with lower error values. {The computational superiority of the proposed model over simulation-based approaches allows one to calculate channel responses in much smaller times. For comparison, in simulations with $N=\SI{e6}{}$ molecules, $\Delta t = \SI{e-4}{\second}$, and $D=\SI[per-mode=symbol]{79.4}{\micro\meter\squared \per \second}$, $3$ spatial coordinates of each molecule is updated at each time step. Also, each molecule is checked whether it is absorbed. Therefore the computational complexity is roughly $\mathcal{O}\left(8 \times 10 \times N \times \frac{t}{\Delta t} \right) = \mathcal{O}\left(\SI{8e11}{}t \right) \sim \mathcal{O}\left(\SI{e12}{}t \right)$.}

On the other hand, the recursive model can be analytically calculated with a convolution and subtraction operation. To the best of our knowledge, the complexity of convolution is $\mathcal{O}\left( n^2 \right) = \mathcal{O}\left( \frac{t}{\Delta t}^2 \right)$. Additionally, $n=\frac{t}{\Delta t}$ subtraction operations are conducted at each time step, which makes the computational complexity $\mathcal{O}\left(\SI{e8}{} \left(t^2 + \SI{e-4}{}t\right) \right) \sim \mathcal{O}\left(\SI{e8}{} t^2 \right)$. Since the channel characteristics saturate quickly in time depending on the spatial parameters, the recursive model has a lower complexity.

Considering previously absorbed molecules, the recursive model predicts the channel behavior by removing the further effects of those molecules. It however, considers the higher-degree effects of those molecules, such as the impacts of the molecules absorbed by an Rx, and then released again and absorbed by the other Rx. In cases where the receivers are closely-positioned, those higher-degree impacts may have a significant effect on especially closely-packed topologies. Considering the increased complexity required, an simplified modeling method is also proposed. In this model, overlapping effects of the absorbed molecules are neglected, and recursive expressions are reduced to explicit analytical expressions. In this way, much simpler expressions for molecular SIMO topologies are acquired, and the channel response is modeled via a single expression with reduced complexity.

The correlation between the recursive model and the simplified model is outstanding, as the channel response models match significantly. As aforementioned before, the closer the receivers, the higher the impact of the higher-degree effects of absorbed molecules is. {For smaller separation angles that are close to $0$, these neglected terms may have a significant effect, hence the slight deviation between the responses of the recursive model and the simplified model.} For larger separation angles, those terms have nearly zero effect, and the recursive model and the simplified model responses match promisingly. Furthermore, as the recursive model fits quite well with the computer simulation data, it is shown that the simplified model fits with the computer simulation data, as well.

Lastly, this study focuses on providing a proper and relatively simple analytical model for molecular SIMO topologies, since it neglects the exact space coordinates of each molecule absorbed, which would provide a more accurate model. Further improvements on release point assumptions, small separation angle deviations and shadowing phenomenon are left as future work.

\appendix

Here, the region of convergence for analytical derivation of the recursive model is derived. In order to expand $\frac{1}{1-c \exp\left(-k \sqrt{s} \right)}$ into an infinite sum, we need to state
\begin{equation}
    \vert c \exp\left(-k \sqrt{s}\right) \vert < 1.
\end{equation}
Let $-k\sqrt{s}$ be a complex number in form of $x+yi$, where $\Re (-k\sqrt{s}) = x$ and $\Im(-k\sqrt{s}) = y$. Then,
\begin{equation}
\vert c \exp\left(-k \sqrt{s}\right) \vert = \vert c \exp(x)\exp(yi) \vert.
\end{equation}
The expression $c \exp(x)\exp(yi)$ can be separated as the magnitude part of $c \exp(x)$, and the phase angle part of $\exp(yi)$. The inequality then becomes
\begin{equation}
\vert c \exp(x) \vert < 1 \rightarrow c  < \exp(-x).
\end{equation}
Recalling the equality
\begin{equation}
-k\sqrt{s} = x + yi \rightarrow \sqrt{s} = \frac{x}{-k} + \frac{yi}{-k},
\end{equation}
we can let $s = \sigma + i\omega$ to obtain
\begin{equation}
\sqrt{\sigma + i\omega} = \frac{x}{-k} + \frac{yi}{-k}.
\end{equation}
By applying basic algebra, the equality becomes
\begin{equation}
\frac{x}{-k} = \sqrt{\frac{\sqrt{\sigma^2+\omega^2}+\sigma}{2}}, \quad \frac{y}{-k} = \sqrt{\frac{\sqrt{\sigma^2+\omega^2}-\sigma}{2}},
\end{equation}
and placing the expression into the inequality gives
\begin{equation}
\label{eq:appendix_final}
c < \exp\left( k \sqrt{\frac{\sqrt{\sigma^2+\omega^2}+\sigma}{2}} \right).
\end{equation}
This concludes the region of convergence for infinite sum expansion.

\bibliographystyle{IEEEtran}
\bibliography{references.bib}
\end{document}